\documentclass[twocolumn,numberedappendix,twocolappendix]{aastex631}
\turnoffedit
\usepackage{amsmath, amssymb}
\usepackage{bm}
\usepackage{color}
\usepackage{float}
\usepackage{placeins}
\usepackage{xcolor}
\usepackage[utf8]{inputenc}
\usepackage[T1]{fontenc}
\usepackage{apjfonts}
\usepackage{mathtools}
\usepackage{enumitem}
\usepackage{epstopdf}
\usepackage{graphicx}
\usepackage{listings}
\usepackage[all]{hypcap}
\usepackage[encapsulated]{CJK} 
\usepackage{ucs}
\newcommand{\cntext}[1]{\begin{CJK}{UTF8}{gbsn}#1\end{CJK}}
\usepackage{devanagari} 
\usepackage{wrapfig}

\providecommand{\sorthelp}[1]{} 

\newcommand\wmap{\text{WMAP}}
\newcommand\planck{\text{Planck}}

\newcommand\cg{{\scshape Cosmoglobe}}
\newcommand\wk{\text{W$K$}}
\newcommand\wka{\text{W$Ka$}}
\newcommand\wq{\text{W$Q$}}

\newcommand{\mr}[1]{\mathrm{#1}}
\newcommand{\ms}[1]{\mathsf{#1}}

\newcommand{\GHz}{\,\mathrm{GHz}}

\newcommand{\nside}{N_\mr{side}}
\newcommand{\ttt}[1]{\texttt{#1}}
\newcommand\hp{\ttt{healpy}}
\newcommand\pysm{\ttt{PySM}}

\definecolor{citecolor}{rgb}{0.08,0.30,0.85}

\newcommand{\jhu}{William H. Miller III Department of Physics and Astronomy, Johns Hopkins University, 3701 San Martin Drive, Baltimore, MD 21218, USA}
\newcommand{\villanova}{Department of Physics, Villanova University, 800 Lancaster Avenue, Villanova, PA 19085, USA}
\newcommand{\goddard}{NASA Goddard Space Flight Center, 8800 Greenbelt Road, Greenbelt, MD 20771, USA}
\newcommand{\upenn}{Department of Physics and Astronomy, University of Pennsylvania, 209 South 33rd Street, Philadelphia, PA 19104, USA}
\newcommand{\oslo}{Institute of Theoretical Astrophysics, University of Oslo, P.O. Box 1029 Blindern, N-0315 Oslo, Norway}
\newcommand{\cfa}{Center for Astrophysics, Harvard \& Smithsonian, 60 Garden Street, Cambridge, MA 02138, USA}

\newcommand{\ucsc}{Departamento de Ingenier\'{i}a El\'{e}ctrica, Universidad Cat\'{o}lica de la Sant\'{i}sima Concepci\'{o}n, Alonso de Ribera 2850, Concepci\'{o}n, Chile}

\newcommand{\argonne}{High Energy Physics Division, Argonne National Laboratory, 9700 South Cass Avenue, Lemont, IL 60439, USA}
\newcommand{\uchicago}{Department of Astronomy and Astrophysics, University of Chicago, 5640 South Ellis Avenue, Chicago, IL 60637, USA}
\newcommand{\MIT}{MIT Kavli Institute, Massachusetts Institute of Technology, 77 Massachusetts Avenue, Cambridge, MA 02139, USA}

\shortauthors{Shi et al.}

\begin{document}
\begin{CJK*}{UTF8}{gbsn}

\title{Sensitivity-Improved Polarization Maps at 40 GHz with CLASS and WMAP data}

\correspondingauthor{Rui Shi}
\email{rshi9@jhu.edu}
\author[0000-0001-7458-6946]{Rui Shi (时瑞)}
\affiliation{\jhu}
\author[0000-0002-8412-630X]{John W.~Appel}\affiliation{\jhu}
\author[0000-0001-8839-7206]{Charles L.~Bennett}\affiliation{\jhu}
\author[0000-0001-8468-9391]{Ricardo Bustos}\affiliation{\ucsc}
\author[0000-0003-0016-0533]{David T.~Chuss}
\affiliation{\villanova}
\author[0000-0002-1708-5464]{Sumit Dahal}\affiliation{\goddard}
\author[0000-0002-0552-3754]{Jullianna Denes~Couto}\affiliation{\jhu}
\author[0000-0001-6976-180X]{Joseph R.~Eimer}\affiliation{\jhu}
\author[0000-0002-4782-3851]{Thomas~Essinger-Hileman}\affiliation{\goddard}
\author[0000-0003-1248-9563]{Kathleen Harrington}\affiliation{\argonne}\affiliation{\uchicago}
\author[0000-0001-7466-0317]{Jeffrey Iuliano}\affiliation{\upenn}\affiliation{\jhu}
\author[0000-0002-4820-1122]{Yunyang Li (李云炀)}
\author[0000-0003-4496-6520]{Tobias A.~Marriage}\affiliation{\jhu}
\author[0000-0002-4436-4215]{Matthew A.~Petroff}\affiliation{\cfa}
\author[0000-0003-4189-0700]{Karwan Rostem}\affiliation{\goddard}
\author[0009-0005-6067-612X]{Zeya Song (宋泽雅)}\affiliation{\jhu}
\author[0000-0003-3487-2811]{Deniz A. N. Valle}\affiliation{\jhu}
\author[0000-0002-5437-6121]{Duncan J.~Watts}\affiliation{\oslo}
\author[0000-0003-3017-3474]{Janet L.~Weiland}\affiliation{\jhu}
\author[0000-0002-7567-4451]{Edward J.~Wollack}\affiliation{\goddard}
\author[0000-0001-5112-2567]{Zhilei Xu (\cntext{徐智磊}\!\!)}\affiliation{\MIT}

\begin{abstract}
Improved polarization measurements at frequencies below $70~\GHz$ with degree-level angular resolution are crucial for advancing our understanding of the Galactic synchrotron radiation and the potential polarized anomalous microwave emission and ultimately benefiting the detection of primordial $B$ modes.
In this study, we present sensitivity-improved $40~\GHz$ polarization maps obtained by combining the CLASS $40~\GHz$ and WMAP $Q$-band data through a weighted average in the harmonic domain.
The decision to include WMAP $Q$-band data stems from similarities in the bandpasses.
Leveraging the accurate large-scale measurements from WMAP $Q$ band and the high-sensitivity information from CLASS 40 GHz band at intermediate scales, the noise level at $\ell\in[30, 100]$ is reduced by a factor of $2-3$ in the map space.
A pixel domain analysis of the polarized synchrotron spectral index ($\beta_s$) using WMAP $K$ band and the combined maps (mean and 16/84th percentile across the $\beta_s$ map: $-3.08_{-0.20}^{+0.20}$) reveals a stronger preference for spatial variation (PTE for a uniform $\beta_s$ hypothesis smaller than 0.001) than the results obtained using WMAP $K$ and $Ka$ bands ($-3.08_{-0.14}^{+0.14}$).
The cross-power spectra of the combined maps follow the same trend as other low-frequency data, and validation through simulations indicates negligible bias introduced by the combination method (sub-percent level in the power spectra).
The products of this work are publicly available on \ttt{LAMBDA}\footnote{\href{https://lambda.gsfc.nasa.gov/product/class/class_prod_table.html}{https://lambda.gsfc.nasa.gov/product/class/class\_prod\_table.html}}.
\end{abstract}

\keywords{
    \href{http://astrothesaurus.org/uat/322}{Cosmic microwave background radiation (322)};  
    \href{http://astrothesaurus.org/uat/1146}{Observational Cosmology (1146)}; 
    \href{http://astrothesaurus.org/uat/1858}{Astronomy Data Analysis (1858)}
}

\section{Introduction}\label{sec:intro}
The polarized microwave sky has been measured with increasing precision and angular resolution over the past three decades by two all-sky space experiments: the Wilkinson Microwave Anisotropy Probe (\wmap) \citep{hinshaw2012, bennett2012} and the \planck{} satellite \citep{planck2016-l04, planck2016-l06}.
Ground-based experiments also made significant contributions to the field, e.g., 
the Atacama B-mode Search experiment \citep{simon2016characterizing, kusaka2018results}, 
the Atacama Cosmology Telescope (ACT, \citealt{li2021situ, qu2024atacama}),
the BICEP/\textit{Keck} \textit{Array} program \citep{hui2018bicep, BK-XV20}, 
the Cosmology Large Angular Scale Surveyor (CLASS, \citealt{essinger-hileman14spie, eimer23}),
the {\scshape Polarbear} experiment (\citealt{inoue2016polarbear, adachi2022improved}, the predecessor of the Simons Array experiment, \citealt{stebor2016simons}),
the Q-U-I JOint Tenerife Experiment (QUIJOTE, \citealt{rubino2010quijote, quijoteIV}), 
and the South Pole Telescope (SPT, \citealt{chown2018maps, sobrin2022design}).
Despite challenges posed by low-frequency $1/f$ variations associated with atmospheric, instrumental, and calibration, ground-based experiments have excelled in measuring intermediate to small angular scales by employing high throughput optical systems suitably stabilized to enable long-duration observations with high sensitivity.

\begin{figure*}[!t]
    \centering
    \begin{minipage}{0.45\linewidth}
        \centering
        \raisebox{-0.5\height}{\includegraphics[width=\linewidth]{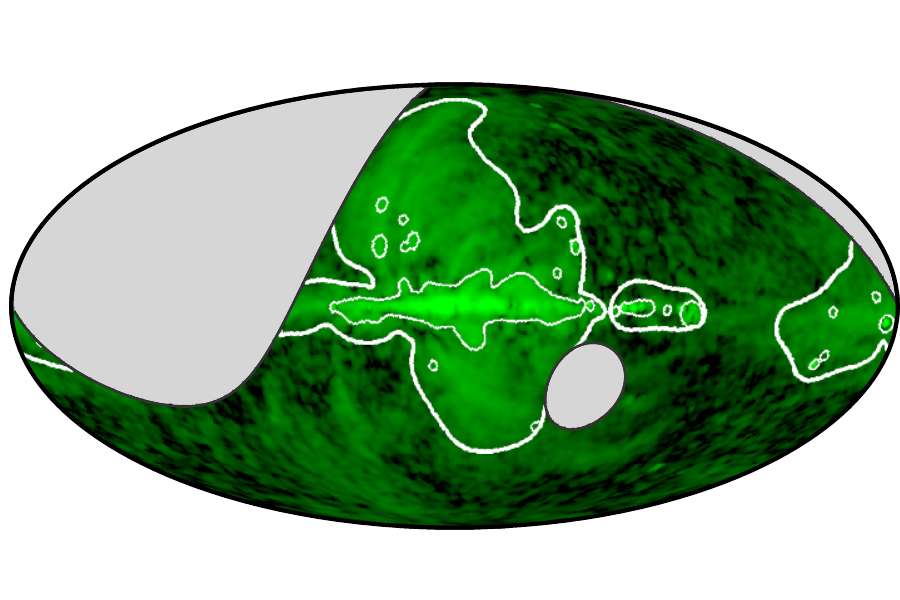}}
    \end{minipage}
    \hfill
    \begin{minipage}{0.52\linewidth}
        \centering
        \raisebox{-0.5\height}{\includegraphics[width=\linewidth]{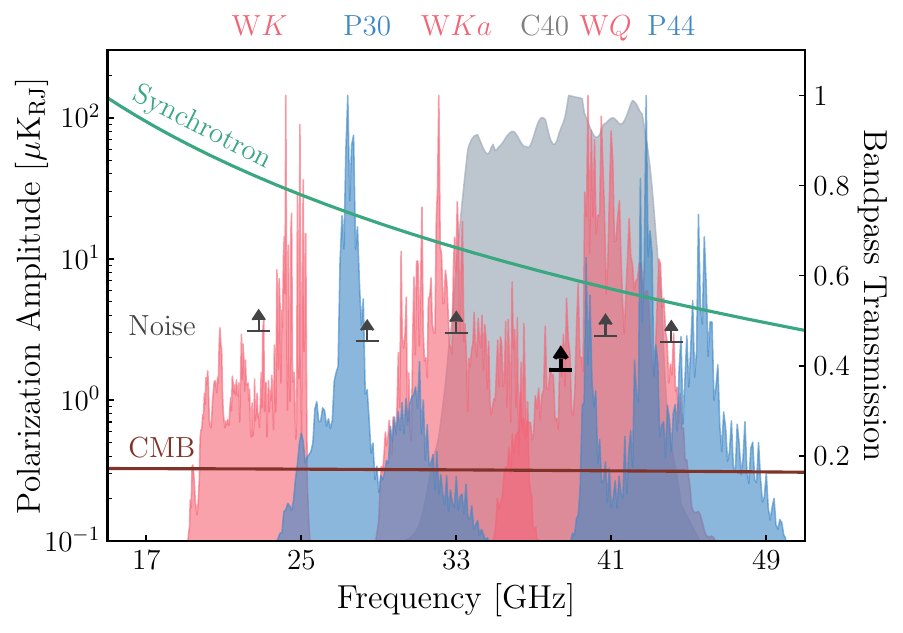}}
    \end{minipage}
    \caption{Synchrotron Survey Overview. 
    \textit{Left}: the green scale in the map shows the polarization intensity of the Planck \ttt{Commander} synchrotron map at $30~\GHz$ \citep{planck2016-l04}. 
    Unless explicitly stated, all maps presented in this paper use the Mollweide projection and are in the Galactic coordinate system.
    The gray regions are beyond the CLASS survey boundary.
    The CLASS \ttt{s9} mask \citep{eimer23} is outlined by thicker white curves, with thinner curves delineating CLASS \ttt{s0} and bright source mask (only the part inside the \ttt{s9} mask region).
    The region in between the thinner and thicker curves captures most of the polarized synchrotron power while excluding the brightest regions.
    \textit{Right}: the green curve shows the polarized synchrotron amplitude in Rayleigh-Jeans temperature units with power-law ($\beta_s=-3.1$ power law) frequency dependence.
    The amplitude at $30~\GHz$ is obtained as the mean of polarization amplitude between thinner and thicker borders in the left panel.
    The red curve represents the polarized CMB amplitude, and the horizontal bars with up arrows are the noise levels for frequency bands listed above the figure.
    The bolded arrow indicates the noise level of the combined maps produced in this work.
    These are obtained as the mean polarization amplitude from the 200 CMB or noise simulations in the same region.
    Data and simulations were smoothed to $\mathrm{FWHM}=2^\circ$ Gaussian beam for this calculation.
    The shaded regions are the bandpasses for different bands, normalized to unity at maximum.
    }
    \label{fig:background}
\end{figure*}

This paper aims to combine the recently released CLASS 40 GHz band data \citep{eimer23} with the WMAP $Q$-band data \citep{bennett2012} through a weighted average in the harmonic domain.
The CLASS $40~\GHz$ band data achieved higher sensitivity than the analogous frequencies from satellite measurements in the range $10 < \ell < 100$, while the largest scale signal had been suppressed by the filters applied in the time-ordered data processing \citep{Li23}.
We used the WMAP $Q$-band data to compensate for the missing power at the largest angular scales.
The decision to use the WMAP $Q$ band over other low-frequency (such as \planck{} $44~\GHz$ band) measurements was motivated by the better match between the bandpasses of CLASS $40~\GHz$ and WMAP $Q$ band (Figure \ref{fig:background}).
The combined products take advantage of accurate large angular scale information from the WMAP $Q$ band and precise measurements from the CLASS 40 GHz band at intermediate angular scales. 

The polarized sky signal at 40 GHz is dominated by the Galactic synchrotron radiation at scales larger than degree-level, with contributions from CMB, Galactic thermal dust radiation, and potentially anomalous microwave emission (AME) \citep{planck2016-l04,hoz2023quijote,svalheim2023beyondplanck,watts2023cosmoglobe,genova2017quijote,herman2023beyondplanck}.
Improved measurements at $40~\GHz$ spanning scales larger than degree level are promising for advancing our understanding of the spatial variation and any steepening or flattening of the polarized synchrotron spectral index, as well as the polarization fraction (currently there are only upper limits) of the anomalous microwave emission \citep{abazajian2016cmbs4,hensley2022simons}.
These improvements will be crucial for the potential future detection of the primordial $B$ modes \citep{ade2019simons,wolz2023simons,litebird2023}.

Analogous combinations have been done for SPT \citep{crawford2016maps, chown2018maps} and ACT \citep{aghanim2019pact, madhavacheril2020atacama, naess2020atacama} data at higher frequencies (>90 GHz).
A joint analysis of \wmap{} and \planck{} data has been conducted in \citep{watts2023cosmoglobe}.
The data products have reinforced our comprehension of CMB lensing \citep{omori2019dark, omori2023joint, qu2024atacama}, cosmic infrared background \citep{kilerci2023infrared}, Sunyaev Zel'dovich effect \citep{calafut2021atacama,gatti2022cross,radiconi2022thermal,mallaby2023kinematic,meinke2023evidence}, Galactic dust science \citep{guan2021atacama,lowe2022study,dorcova2024atacama}, and related fields.

The structure of the paper is as follows: in Section \ref{sec:data} we introduce the data and simulations, and in Section \ref{sec:method} we describe the method to combine the data. 
We present the bandpasses, combined maps, and noise power spectra in Section \ref{sec:result}, and a pixel-based polarized synchrotron index analysis in Section \ref{sec:beta_s}. 
We summarize in Section \ref{sec:conclusion}.

\begin{deluxetable*}{lcccccc}[t]
\tablecaption{The abbreviations used in this paper, approximate beam sizes, reference frequencies ($\nu_\mathrm{ref}$), temperature-to-antenna (T-to-A) conversion factors (or equivalently, the conversion between thermodynamic temperature $T_\mr{CMB}$ and Rayleigh-Jeans temperature $T_\mr{RJ}$), map level scaling factors to convert to $40~\GHz$, and the dust template scaling factors for different bands.\label{tab:all_data}}
\tablehead{
\colhead{} & \colhead{Abbreviation} & \colhead{FWHM$^a$ [$\mr{arcmin}$]} & \colhead{$\nu_\mathrm{ref}$ [$\GHz$]} & \colhead{T-to-A$^b$} & \colhead{Scale to 40 GHz$^c$} & \colhead{Dust template scaling$^d$}
}
\startdata
WMAP $K$ & \wk & $52.8$ & $22.8$ & $0.9370$ & $0.1751$ & $0.0016$\\
Planck 30 GHz & P30 & $32.4$ & $28.4$ & $0.9776$ & $0.3459$ & $/^e$\\
WMAP $Ka$ & \wka & $39.6$ & $33$ & $0.9416$ & $0.5508$ & $0.0029$\\
CLASS 40 GHz & C40 & $93.6$ & $38$ & $0.9310$ & $0.8530$ & $0.0037$\\
WMAP $Q$ & \wq & $30.6$ & $40.7$ & $0.9311$ & $1.0553$ & $0.0040$\\
Planck 44 GHz & P44 & $27$ & $44.1$ & $0.9380$ & $1.3532$ & $/$\\
Combined & CW & $60$ & $40$ & $--^f$ & $/$ & $/$\\
(r)Planck 353 GHz$^g$ & (r)P353 & $5$ ($93.6$)$^g$ & 353 & $/$ & $/$ & $/$
\enddata
\tablecomments{~\\
$^a$The FWHM of the Gaussian approximation of the beam profile. We used the beam profiles when processing the data unless explicitly mentioned: CLASS and \wmap{} beam profiles are available on \ttt{LAMBDA}, and those for \planck{} are in their reduced instrument model (\ttt{LFI\_RIMO\_R3.31.fits}).\\
$^b$T-to-A conversion factors for synchrotron spectrum were calculated assuming a $\beta_s=-3.1$ power law \citep{planck2016-l04}. The color correction factors were computed following the method in \citet{eimer23} for \wmap{} and CLASS data, and the public code \texttt{fastcc} \citep{fastcc} for \planck.\\
$^c$The factors to scale maps (antenna temperature) to $40~\GHz$ assuming a $\beta_s=-3.1$ power law.\\
$^d$The factors to scale the polarized thermal dust template (original or reobserved \planck{} 353 GHz data) following a $\beta_d=1.53$ and $T_\mr{dust}=19.6~\mr K$ modified blackbody spectrum \citep{planck2016-l04}. The factors concern three processes: the T-to-A conversion for \planck{} $353~\GHz$ band, the modified-blackbody scaling, and the A-to-T conversion at the corresponding bands.\\
$^e$Not applicable / Unused.\\
$^f$The conversion factor for the combined data relies on spherical harmonic degree $\ell$ and order $m$; see discussion in Section \ref{ssec:bandpass}.\\
$^g$The rPlanck 353 GHz refers to the reobserved Planck 353 GHz data processed with the CLASS $40~\GHz$ data pipeline \citep{Li23}. The beam size provided in parentheses corresponds to the value for rP353.
}
\end{deluxetable*}
\section{Data and Simulations}\label{sec:data}
In this section, we briefly introduce the experiments and data used in this work.
A summary of relevant properties can be found in Figure \ref{fig:background} and Table \ref{tab:all_data}.

\subsection{CLASS}\label{ssec:class}
The CLASS telescope array is located on Cerro Toco in the Atacama Desert of northern Chile (longitude $67^\circ$W, latitude $23^\circ$S) and surveys the sky with single-frequency-band telescopes centered at 40 GHz and 90 GHz, as well as a dual-band 150/220 GHz telescope \citep{essinger-hileman14spie, harrington16spie}. 
During typical CMB scans conducted until May 2022, the CLASS telescopes observed the sky at a constant elevation angle of $45^\circ$, scanning azimuthally at $1$ or $2~\mr{deg/s}$.
Additionally, their boresight angles change daily between $-45^\circ$ and $+45^\circ$ with increments of $15^\circ$.
Adopting this scanning strategy and possessing a large field-of-view ($\sim20^\circ$ in azimuth and $15^\circ$ in elevation, \citealt{eimer12spie, xu20}), the CLASS telescopes effectively map $\sim 75\%$ of the sky daily, covering declination angles ($\delta$) ranging from $-76^\circ$ to $30^\circ$.
The front-end polarization modulator -- variable-delayed polarization modulator (VPM, \citealt{chus12vpm}) -- on the CLASS telescopes significantly improves the stability of the observation \citep{harrington21,cleary22}, enabling stable measurements even for the largest angular scales ($\ell<20$).

In this paper, we used the fully coadded 40 GHz maps and the two base splits comprising observations made from 31 August 2016 to 19 May 2022 \citep{eimer23}.
Details on the CLASS $40~\GHz$ data pipeline can be found in \citet{Li23}.
The reference frequency for these maps is $38~\GHz$, and the telescope beam can be approximated with a Gaussian beam with FWHM $1.56^\circ$.
We also used the 200 noise simulations of the total map and the noise simulations of the two splits (600 simulations in total).
In the polarized synchrotron spectral index analysis (Section \ref{sec:beta_s}), we used the reobserved \planck{} (rPlanck) PR4 353 GHz maps.
The Stokes $Q$/$U$ components in the reobserved maps were created by first projecting the original maps into the time domain assuming the CLASS $40~\GHz$ pointing model, and then processing the time streams following the same methodology applied to the CLASS $40~\mathrm{GHz}$ demodulated data \citep{Li23}.
The rPlanck PR4 353 GHz maps were used as the polarized thermal dust template for the CLASS $40~\GHz$-band data, as they underwent identical filtering processes.
The absolute calibration for CLASS $40~\GHz$ maps was determined by comparing the CLASS data to the bright diffuse synchrotron signal in reobserved WMAP $Ka$ and $Q$ bands \citep{eimer23}. 
The calibration uncertainty is 5\%, primarily due to the 0.5 GHz uncertainty on the band center of CLASS $40~\GHz$ \citep{dahal22}.

All the CLASS data products (including the rPlanck maps) used in this paper are available on \ttt{LAMBDA}.\footnote{\href{https://lambda.gsfc.nasa.gov/product/class/class_prod_table.html}{https://lambda.gsfc.nasa.gov/product/class/class\_prod\_table.html}}

\subsection{WMAP}\label{ssec:wmap}
The WMAP satellite \citep{bennett2003a} conducted a 9-year all-sky survey from 2001 to 2009 \citep{hinshaw2012, bennett2012}.
It observed the sky at five frequency bands: $K$ ($23~\GHz$), $Ka$ ($33~\GHz$), $Q$ ($41~\GHz$), $V$ ($61~\GHz$), and $W$ ($94~\GHz$). 

In this paper, we used the 9-year maps of $K$, $Ka$, and $Q$ bands; the reference frequencies and approximate beam sizes can be found in Table \ref{tab:all_data}.
We made two splits for each band by coadding single-year maps (year 1--4 and year 5--9).
We preprocess the maps as follows:
\begin{enumerate}[itemsep=-3pt]
    \item Transform the maps into the Celestial coordinate system.
    \item Smooth the maps from the original telescope beams to an FWHM=$1^\circ$ Gaussian beam profile.
    \item Downgrade the maps to \ttt{HEALPix} resolution $\nside$=256.
    \item Apply the CLASS survey boundary mask $\delta\in(-76^\circ, 30^\circ)$.
\end{enumerate}

Full-sky noise simulations in this work were generated using the noise covariance matrices, following the method described in \citet{larson2011seven}.
To capture the correlation at large angular scales, we first created $\nside=16$ simulations by sampling according to the full noise covariance matrices and then upgraded them to $\nside=256$.
We created white noise realizations at $\nside=256$ by sampling according to the per-pixel $Q$/$U$ covariance matrices. 
Finally, we added the white noise realizations (with mean within each $\nside=16$ pixel removed) with the noise simulations created at $N_\mr{nside}=16$.
We made 200 noise simulations for each 9-year map and its two single-year coadded splits.
They were preprocessed in the same way as the data.
The \wmap{} absolute calibration source was the dipole anisotropy induced by \wmap{}'s motion with respect to the CMB rest frame \citep{hinshaw2009new, jarosik11}. The calibration uncertainty for \wmap{} is 0.2\% \citep{bennett2012}.

All the \wmap{} data products used in this paper are available on \ttt{LAMBDA}.\footnote{\href{https://lambda.gsfc.nasa.gov/product/wmap/dr5/m_products.html}{https://lambda.gsfc.nasa.gov/product/wmap/dr5/m\_products.html}}

\subsection{Planck}\label{ssec:planck}
The \planck{} satellite \citep{planck2016-l01} carried out a high-precision all-sky survey mission between 2009 and 2013, observing the sky in nine frequency bands: 30, 44, and 70 by the Low Frequency Instrument (LFI, \citealt{planck2016-l02}), 100, 143, 217, 353, 545, and 857 GHz by the High Frequency Instrument (HFI, \citealt{planck2016-l03}).

In this study, we used the PR4 bandpass-corrected 353 GHz maps as the polarized thermal dust template in Section \ref{sec:beta_s}.
We used the PR3 full map and half-ring splits of Planck 30 GHz and 44 GHz bands for validation (Section \ref{ssec:maps}, Appendix \ref{sec:dataspectra}). 
The reference frequencies and approximate beam sizes can be found in Table \ref{tab:all_data}.
There are 300 Full Focal Plane (FFP 10) noise simulations available for each full and half-ring split map, and we used the first 200 with indices ranging from 0 to 199.
These maps and noise simulations were preprocessed in the same way as the \wmap{} data (Section \ref{ssec:wmap}).
Due to the limited number of (100) simulations available for PR4, we used the PR3 products for the LFI data.
The \planck{} data employed the same absolute calibration source as the \wmap{} data. The calibration uncertainty for PR3 LFI maps is $0.07-0.11\%$ \citep{planck2016-l02}, while for PR4 $353~\GHz$ maps, the uncertainty is $0.05\%$ \citep{planck2020-LVII}.

The Planck data and simulations are available on Planck Legacy Archive.\footnote{\href{https://pla.esac.esa.int}{https://pla.esac.esa.int}}

\subsection{Additional simulations}\label{ssec:sims}
We used 200 CMB simulations in this work.
They were generated according to the Planck 2018 best-fit CMB power spectra,\footnote{\ttt{COM\_PowerSpect\_CMB-base-plikHM-TTTEEE-lowl-lowE-\\lensing-minimum-theory\_R3.01.txt}} originally made at $\nside=256$ in Galactic coordinates and preprocessed identically to that of the \wmap{} data (Section \ref{ssec:wmap}).
\ttt{PySM} \citep{pysm} simulations were used to validate the combination pipeline (Section \ref{sec:method}, Appendix \ref{sec:verification}) and the polarized synchrotron spectral index analysis (Section \ref{sec:beta_s}, Appendix \ref{sec:beta_s_verify}).
We also generated reobserved simulations by processing all simulations described in this section with the CLASS $40~\GHz$ data pipeline \citep{Li23}.

\section{Harmonic domain combination}\label{sec:method}
In this section, we introduce the combination method following the approach of \citet{chown2018maps}.
We begin with the formalism (Section \ref{ssec:notation}), followed by an explanation of the combination method in Section \ref{ssec:wave}.
We then describe the filtering matrix in Section \ref{ssec:filtermatrix} and the harmonic domain noise in Section \ref{ssec:noise}, and summarize the products we made in Section \ref{ssec:summary}.

\subsection{Formalism}\label{ssec:notation}
The spherical harmonic coefficients for linear polarization measurements can be written as:
\begin{equation}
    \begin{aligned}
    a_{2,\ell m}&=\int d\Omega ~(Q+iU)(\hat{\pmb{n}})~_2Y_{\ell m}(\hat{\pmb{n}}),\\
    a_{-2,\ell m}&=\int d\Omega ~(Q-iU)(\hat{\pmb{n}})~_{-2}Y_{\ell m}(\hat{\pmb{n}}),
    \end{aligned}\label{eq:QUharm}
\end{equation}
where $Q/U$ are the maps of the linear polarization Stokes parameters and $_{\pm 2}Y_{\ell m}$ are the spin-weighted harmonics with spin weight $s=\pm2$.
For convenience, we work with $E$/$B$-mode spherical harmonics which are linear combinations of $a_{\pm 2,\ell m}$:
\begin{equation}
    \begin{aligned}
    a^E_{\ell m}&=-\frac{1}{2}(a_{2,\ell m}+a_{-2,\ell m}),\\
    a^B_{\ell m}&=-\frac{1}{2i}(a_{2,\ell m}-a_{-2,\ell m}).
    \end{aligned}\label{eq:EBharm}
\end{equation}

Since only linear operations are involved in Equations \ref{eq:QUharm} and \ref{eq:EBharm}, the definitions above can be expressed for pixelized linear polarization maps with matrix operations as:
\begin{equation}
    \pmb{a}^\mr{sky}=\ms{H}\pmb{m}^\mr{sky},
\end{equation}
where $\pmb{a}\equiv(a^E_{\ell m},a^B_{\ell m})^T$ is a column vector with $2N_h$ elements, and $\pmb{m}\equiv(Q_p,U_p)^T$ is a column vector with $2N_p$ elements.
$N_p$ is the number of pixels in a map and $N_h$ is the number of spherical harmonic coefficients with $m\geq 0$.
For maps at $\nside=256$, $N_p=786432$ and $N_h=295296$.
$\ms{H}$ is a $2N_h\times 2N_p$ matrix concerning the harmonic decomposition and the linear combination in Equations \ref{eq:QUharm} and \ref{eq:EBharm}. 
The superscript $^\mr{sky}$ stands for true full-sky signal.

The observed map $\pmb{m}^\mr{obs}$ is related to the true full-sky signal as $\pmb{m}^\mr{obs}=\ms{F}_p\ms{M}(\ms{B}_p\pmb{m}^\mr{sky}+\pmb{n})$ where $\ms{F}_p$, $\ms{M}$, and $\ms{B}_p$ are the matrix representations of the filtering applied in time-ordered data processing, the mask, and the telescope beam convolution, respectively, and $\pmb{n}$ is the instrumental noise. 
The subscript $_p$ denotes the matrices that are defined in the pixel domain.
Then the observed harmonic coefficients can be expressed as:
\begin{equation}
    \begin{aligned}
    \pmb{a}^\mr{obs}&=\ms{H}\pmb{m}^\mr{obs}\\
    &=\ms{H}\ms F_p\ms{M}(\ms{B}_p\pmb{m}^\mr{sky}+\pmb{n})\\
    &\equiv\ms{F}_h\ms{H}\ms{M}(\ms{B}_p\pmb{m}^\mr{sky}+\pmb{n}),
    \end{aligned}
\end{equation}
where $\ms{H}\ms{F}_p\equiv\ms{F}_h\ms{H}$ and subscript $_h$ denotes the matrix defined in the harmonic domain.

The $\pmb a^\mathrm{obs}$ is different from the $\pmb a^\mathrm{sky}$ even without any significant smoothing and filtering in the time-ordered data processing (i.e., when $\ms B_p$, $\ms F_p$ and $\ms F_h$ can be approximated with identity matrices) because the applied mask (CLASS survey boundary in this case) induces mixing between $E$ and $B$ modes, and between $\ell$'s.
We highlight that the goal of the paper is not to recover $\pmb a^\mathrm{sky}$, but a smoothed, masked map: $\ms M(\ms B_p\pmb m^\mathrm{sky}+\pmb{n})$.

Practically, $\ms{B}_p$ (and its corresponding form in the harmonic domain, $\ms{B}_h$) is realized by the \hp{} function \ttt{smoothalm}, $\ms{M}$ is realized by setting masked pixels to zero, and $\ms{H}$ is realized by the \hp{} function \ttt{map2alm}.
We show how the filtering matrix is approximated in Section \ref{ssec:filtermatrix}.
For brevity, we use $\tilde x$ to denote the observed quantity $x^\mathrm{obs}$ in subsequent sections.

\begin{figure}
    \centering
    \includegraphics[width=\linewidth]{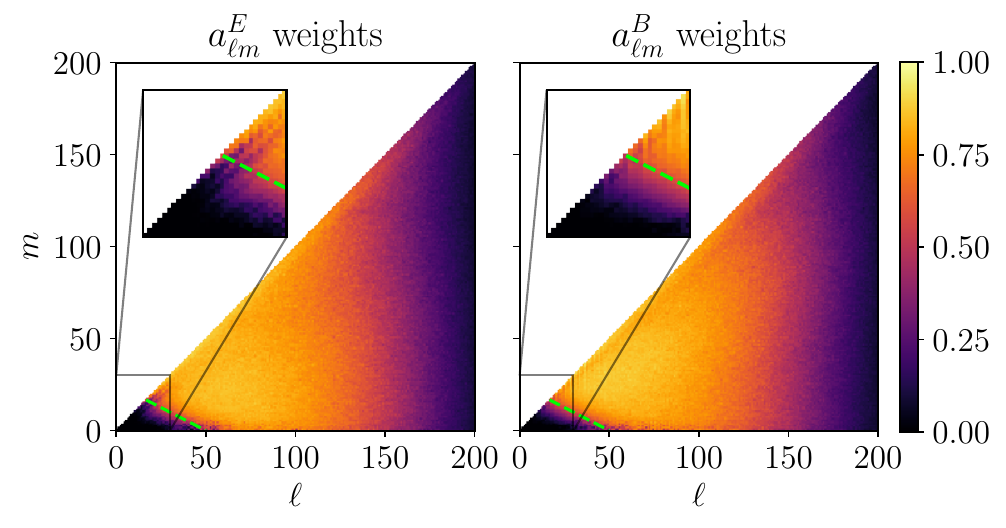}
    \caption{The weights (Equation \ref{eq:weights}) for C40 at each $\ell$ and $m$.
    The weights for $a_{\ell, -m}$ ($m\geq 0$) are equivalent to those for $a_{\ell ,m}$.
    The left panel shows the $E$-mode weights and the right for the $B$-mode. 
    The zoom-in panels show details where $\ell,m<30$.
    The green dashed lines mark where the high-pass filter suppresses the weights by $1/\sqrt 2$.
    We only show the weights at $\ell,m\leq200$, but the full weights extend to $\ell_\mr{max},m_\mr{max}=383$ (beyond which only the \wq{} data contribute in the combined maps).
    At the largest angular scales ($\ell<20$) the weights are small because the high-pass filter and the C40 transfer function suppressed a substantial amount of power.
    At intermediate scales ($\ell\in[20, 150]$) C40 dominates over \wq{} because the noise level is significantly lower.
    At the smallest scales $(\ell>150$) the weights fall mainly because the C40 beam size ($\mathrm{FWHM}\sim1.56^\circ$, corresponds to $\ell=115$) is larger than \wq{}.}
    \label{fig:weights}
\end{figure}
\subsection{Combination method}\label{ssec:wave}
We generated combined maps at $\nside=256$ with a smoothing scale of FWHM=$1^\circ$. The $\ell_\mr{max}$ and $m_\mr{max}$ of these combined maps were set to 767. 
However, since the official CLASS 40 GHz maps were provided at $\nside=128$ and contained limited information beyond $\ell,m=383$, the $a_{\ell m}$'s of the combined maps with $\ell,m>383$ were chosen to be the corresponding modes from WMAP $Q$-band data only.

At $\ell,m\leq383$, we performed a weighted average of the CLASS 40 GHz and WMAP $Q$-band data.
We first processed the C40 data to remove filtering-induced bias, correct the pixel window functions, and match the smoothing scale:
\begin{equation}
    \tilde{\pmb{a}}^\mr{C40,out}=\ms{B}_h^\mr{out}\left(\ms B_h^\mr{C40}\right)^{-1}\ms{P}_h^{256}\left(\ms{P}_h^{128}\right)^{-1}\left(\ms F_h^\mr{C40}\right)^{-1}\tilde{\pmb{a}}^\mr{C40},
\end{equation}
where $\tilde{\pmb{a}}^\mr{C40}=\ms H\tilde{\pmb{m}}^\mr{C40}$ are the spherical harmonic coefficients of the C40 data ($\tilde{\pmb{m}}^\mr{C40}$), and $\ms{P}_h$ is matrix representation of pixel window function in the harmonic domain, which can be realized by the \ttt{healpy} function \ttt{pixwin}.

The combined spherical harmonic coefficients at $\ell,m\leq383$ were made by performing a weighted average of processed CLASS $40~\GHz$ and WMAP $Q$-band data in the harmonic domain:
\begin{equation}
    \tilde{a}_{\ell m}^\mr{CW,out}=w_{\ell m}^\mr{C40} \tilde{a}_{\ell m}^\mr{C40,out} + w_{\ell m}^{\mr WQ} \tilde{a}_{\ell m}^{\mr WQ\mr{,out}}, \label{eq:combine}
\end{equation}
where $\tilde{a}_{\ell m}^{\mr WQ,\mr{out}}$'s are the spherical harmonic coefficients of the WMAP $Q$-band data, and $w_{\ell m}$'s are the harmonic domain inverse noise variance weights:
\begin{equation}
    \begin{aligned}
    w_{\ell m}^\mr{C40}&=g(\ell,m)\frac{1/\big(\sigma_{\ell m}^\mr{C40}\big)^2}{1/\big(\sigma_{\ell m}^\mr{C40}\big)^2+1/\big(\sigma_{\ell m}^{\mr WQ}\big)^2},\\
    w_{\ell m}^{\mr WQ}&=1-w_{\ell m}^\mr{C40},\\
    g(\ell, m)&\equiv\frac{1}{\sqrt{[1+50/(2m+\ell)]^{6}}},\label{eq:weights}
    \end{aligned}
\end{equation}
where $g(\ell,m)$ is a high-pass filter, and the noise level per $(\ell,m)$, $\sigma_{\ell m}$, is introduced in Section \ref{ssec:noise}.
We set weights with values smaller than 0.001 to be zero, and larger than 0.999 to be unity.
We visualize the weights for CLASS $40~\GHz$ data in Figure \ref{fig:weights}.
The form of the high-pass filter was chosen to effectively mitigate the influence of imperfections in the filtering matrix at large scales (Section \ref{ssec:filtermatrix}).
This choice, although rendering the method suboptimal, guarantees that the bias in the final map remains negligible (Appendix \ref{sec:verification}).

Finally, the combined maps can be obtained as:
\begin{equation}
    \tilde{\pmb{m}}^\mr{CW,out}=\ms H^{-1}\tilde{\pmb{a}}^\mr{CW,out},
\end{equation}
and practically, $\ms H^{-1}$ is realized by \hp{} function \ttt{alm2map}.

\subsection{CLASS filtering matrix}\label{ssec:filtermatrix}
\begin{figure}[t]
    \centering
    \includegraphics[width=\linewidth]{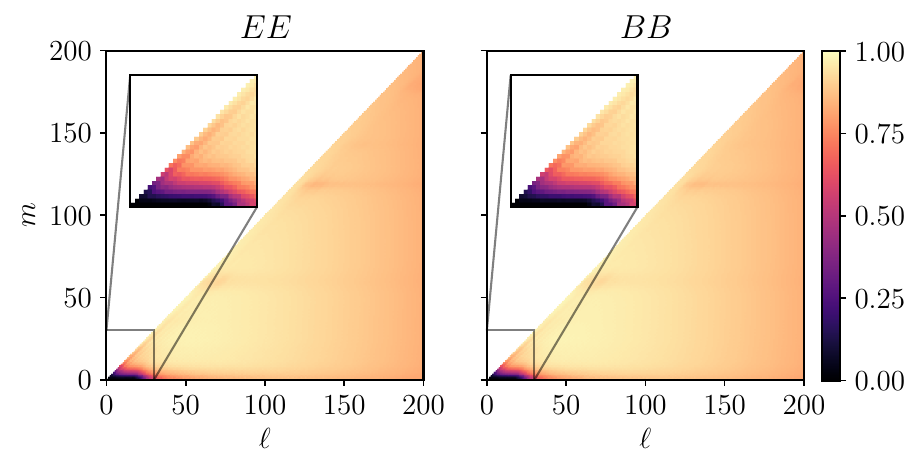}
    \caption{The diagonal elements of the filtering matrix (Equation \ref{eq:tf}) for C40 at each $\ell$ and $m$.
    The element values for $a_{\ell, -m}$ ($m\geq 0$) are equivalent to those for $a_{\ell ,m}$.
    The left panel shows the $EE$ part and the right for the $BB$ part.
    The zoom-in panels show details in $\ell,m<30$.
    We only show the elements at $\ell,m\leq200$, but the full matrix extends to $\ell_\mr{max},m_\mr{max}=383$.
    }
    \label{fig:TFcomp}
\end{figure}

\begin{figure*}
    \centering
    \includegraphics[width=\linewidth]{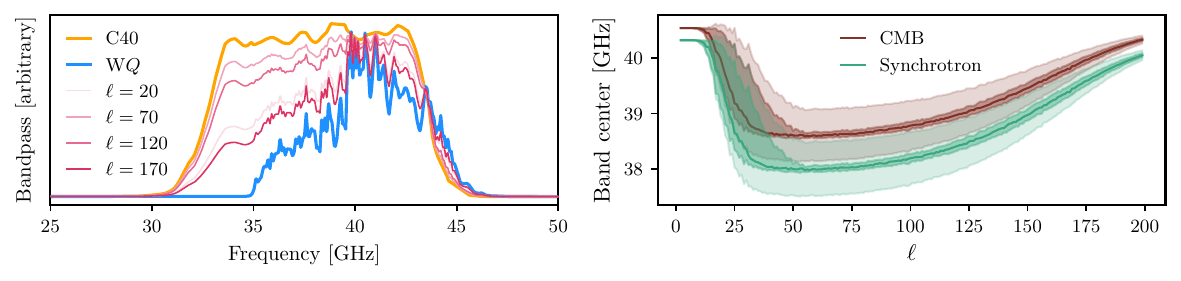}
    \caption{\textit{Left}: the bandpasses for C40 (orange) and \wq{} (blue), and for the combined maps (thinner pink from light to dark) at $\ell=20, 70, 120$ and $170$ ($m=15$). 
    \textit{Right}: the $m$-averaged band center of the combined maps as a function of $\ell$, for CMB (red) and synchrotron (green, assuming $\beta_s=-3.1$) spectrum.
    The darker shaded regions are the 16/84th percentile of the spread of band center per $\ell$, and the lighter shaded regions span the extra uncertainty induced by the $0.5\GHz$ band center uncertainty for C40 data \citep{dahal22}.
    The band centers follow the trend of the weights shown in Figure \ref{fig:weights}.
    The band centers show a $5-7.5\%$ total fractional shift as a function of $\ell$.}
    \label{fig:bandpass}
\end{figure*}

For the CLASS $40~\GHz$ data, the time-ordered data filtering and map-making process are linear operations \citep{Li23}.
Therefore, the connection between the spherical harmonic coefficients of the filtered map ($\pmb{a}^\mr{filt}$) and the unfiltered map ($\pmb{a}^\mr{raw}$) can be expressed as:
\begin{equation}
    \begin{aligned}
    \pmb a^\mr{filt}&=\ms F_h\pmb a^\mr{raw}.
    \end{aligned}
\end{equation}

We approximate the matrix $\ms F_h$ as:
\begin{equation}
    \ms F_{h,ij}\equiv\frac{\left\langle a_i^\mr{filt}\big(a_j^\mr{raw}\big)^*\right\rangle}{\left\langle a_i^\mr{raw}\big(a_j^\mr{raw}\big)^*\right\rangle},\label{eq:tf}
\end{equation}
where $\langle\cdot\rangle$ indicates the ensemble average, $\pmb a^\mr{raw}$ is the spherical harmonic coefficients computed using simulations with flat power spectra in only $EE$ and $BB$ (the $EE$ and $BB$ power are the same), and $\pmb a^\mr{filt}$ are those of the same maps but processed with the CLASS 40 GHz data pipeline \citep{Li23}. 
We used 1000 pairs of such simulations, and the CLASS survey boundary mask was applied before calculating the spherical harmonic coefficients.
The subscripts $_{i, j}$ refer to specific spherical harmonic coefficients.
Since the CLASS 40 GHz maps were provided at $\nside=128$, we choose $\ell_\mathrm{max}=383$, which corresponds to $N_h=73920$ (with $m\geq0$) coefficients for a single mode.
Therefore, the dimension of the full filtering matrix is $147840\times 147840$ for both $E$ and $B$ modes.
We show the diagonal elements of the filtering matrix for C40 in Figure~\ref{fig:TFcomp}.

In this study, we only used the diagonal elements\footnote{The $EB$ and $BE$ relations between $\pmb a^\mr{raw}$'s and $\pmb a^\mr{filt}$'s are dominated by the mixing induced by CLASS survey boundary mask, and are hence ignored in this paper.} of the matrix and dropped the imaginary parts for the calculated filtering matrix because they are negligible.
To estimate the impact of ignoring all off-diagonal components in $\ms F_h$, we defined the fractional error on the filtering-matrix-corrected spherical harmonic coefficients ($\mathsf F_h^{-1}\pmb{a}^\mr{filt}$) as:
\begin{equation}
    \pmb\Delta_h\equiv\sqrt{\frac{\langle|\pmb a_h^\mr{raw}-\ms F_h^{-1}\pmb a_h^\mr{filt}|^2\rangle}{\langle|\ms F_h^{-1}\pmb a_h^\mr{filt}|^2\rangle}},\label{eq:deltah}
\end{equation}
where the numerator is the Root Mean Square (RMS) of the difference between raw and filtering-matrix-corrected filtered spherical harmonic coefficients, and the denominator is the RMS of the filtering-matrix-corrected filtered coefficients.
We computed $\Delta_h$ with the same simulations,
and we used $\Delta_h|(\ms F_h^\mr{C40})^{-1}\tilde{\pmb{a}}^\mr{C40}|$ to represent the expected uncertainty level due to ignoring all off-diagonal components in $\ms F_h$ (Section \ref{ssec:noise}).
Note that the fractional error on any of the harmonic coefficients depends on both the off-diagonal components in $\ms F_h$ and all the other harmonic coefficients, so the $\Delta_h$ in Equation \ref{eq:deltah} only approximates the fractional error of the maps with flat spectra.
The $40~\GHz$ sky spectra are bright and have complicated morphology at the largest angular scales, therefore, we adopted a high-pass filter (Equation \ref{eq:weights}) to compensate for the limitation of this approximation.


\subsection{Harmonic domain noise}\label{ssec:noise}
The noise level per $(\ell, m)$ is computed from 200 noise simulations (Section \ref{sec:data}) and is defined as:
\begin{equation}
    (\pmb\sigma_h)^2\equiv\langle |\pmb n_h\pmb n_h^*|\rangle,\label{eq:noise}
\end{equation}
where $\pmb n_h$'s are vectors formed by the spherical harmonic coefficients ($E$ and $B$ modes) of the noise simulations.
The noise level for the WMAP $Q$ band was calculated exactly as in the equation above.

For the CLASS $40~\GHz$ band, we need two modifications. 
First, we need to apply the filtering matrix correction, which is important for the large angular scale modes.
Second, we need to include the potential fluctuation caused by ignoring all off-diagonal components in the filtering matrix.
The final expression for the C40 noise level per $(\ell, m)$ is:
\begin{equation}
    \begin{aligned}
    \big(\pmb\sigma_h^\mr{C40}\big)^2=&\big(\ms F_h^\mr{C40}\big)^{-1}\langle\pmb n_h^\mr{C40}(\pmb n_h^\mr{C40})^*\rangle\left[(\ms F_h^\mr{C40})^{-1}\right]^T\\
    &+\pmb\Delta_h^2\big|\tilde{\pmb a}^\mr{C40,out}\big|^2,
    \end{aligned}\label{eq:nvar}
\end{equation}
where the superscript $^T$ indicates the transpose of a matrix.
The second term in the equation above gives the expected difference in the spectra power when using a diagonal-only filtering matrix.
The noise simulations were smoothed to FWHM=$1^\circ$ Gaussian beam size and corrected with the pixel window functions before this calculation. 

\subsection{Summary}\label{ssec:summary}
Following the procedure described in this section, we made one total map by combining the fully coadded CLASS $40~\GHz$ map and WMAP $Q$ band 9-year map.
We also created two combined splits: one by combining the C40 m1 base split and \wq\ band year 1-4 coadded split, the other by combining the C40 m2 base split and \wq\ band year 5-9 coadded split.
Furthermore, we created combined noise simulations for the total and the two splits (200 each).
The noise simulations were combined using the same filtering matrix and weights as for the data.
All maps were generated at $\nside=256$ with a smoothing scale of FWHM=$1^\circ$.
In Appendix \ref{sec:dataspectra} and \ref{sec:verification}, we confirm that the filtering matrix and weights utilized for map combination result in negligible bias in the final products.
Maps with higher resolutions can be made upon request.

\section{The Combined Maps}\label{sec:result}
\begin{figure*}[!t]
    \centering
    \includegraphics[width=0.85\linewidth]{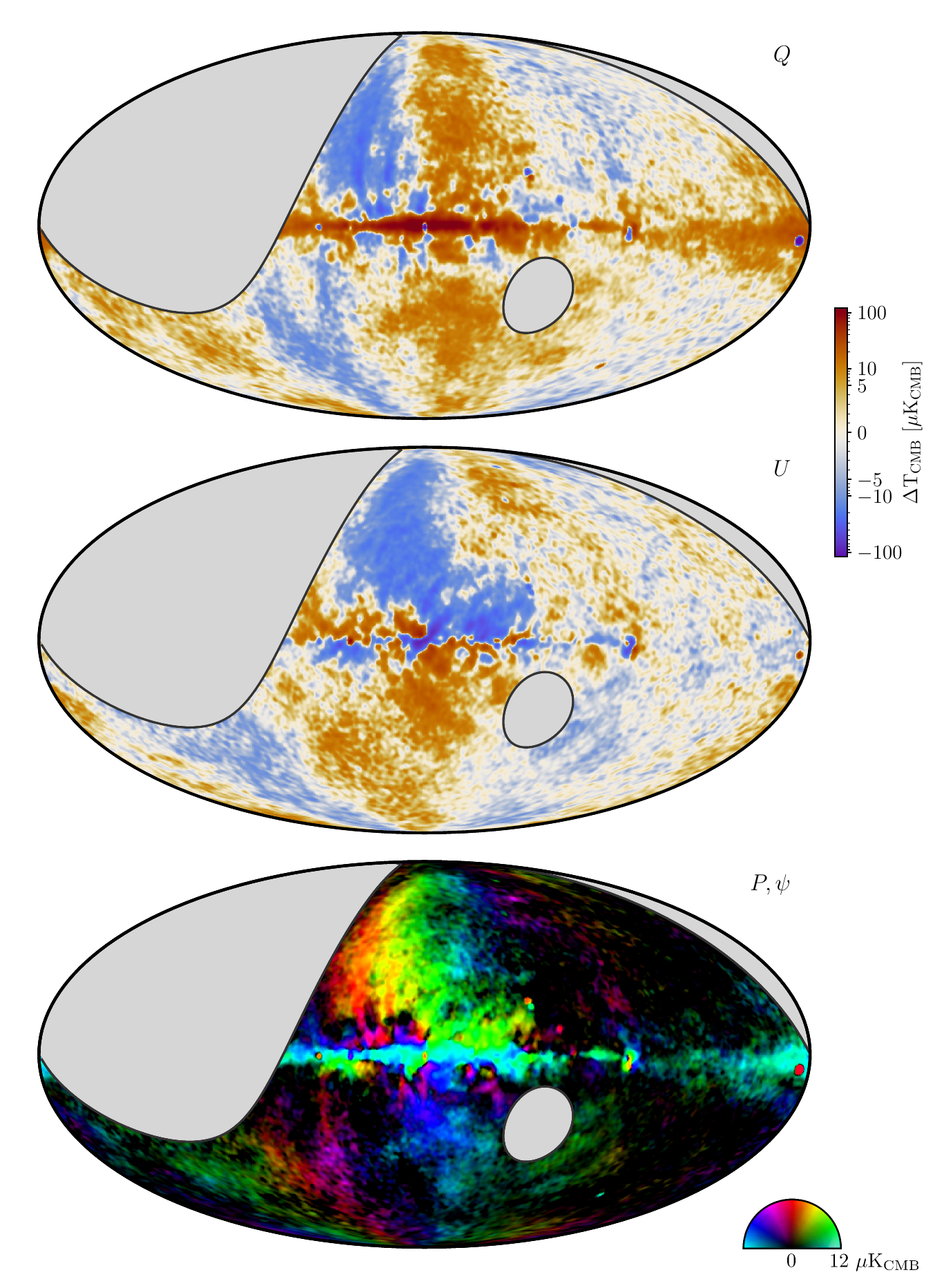}
    \caption{The combined maps.
    \textit{Top} and \textit{Middle}: the Stokes $Q$ and $U$ maps.
    The color scale is linear within $\pm 5~\mathrm{\mu K_\mathrm{CMB}}$ and logarithmic beyond this range.
    \textit{Bottom}: the polarization intensity/angle map (Equation \ref{eq:ppsi}).
    The color scale is visualized on a polar plot: the radial coordinate indicates the polarization intensity (saturates at $12~\mr{\mu K_{CMB}}$), and the angular coordinate refers to the polarization angle. 
    Following the notation used in \citet{planck2014-a12, planck2014-a31}, cyan, yellow, red, and purple colors indicate that the polarization angles are rotated by $(\pm)90^\circ$, $-45^\circ$, $0^\circ$ and $+45^\circ$ with respect to the local meridian, respectively.
    The gray regions are beyond the CLASS survey boundary.
    All maps are smoothed to have an FWHM=$2^\circ$ Gaussian beam size for visualization purposes.
    }
    \label{fig:maps}
\end{figure*}

\begin{figure*}[!t]
    \centering
    \includegraphics[width=\linewidth]{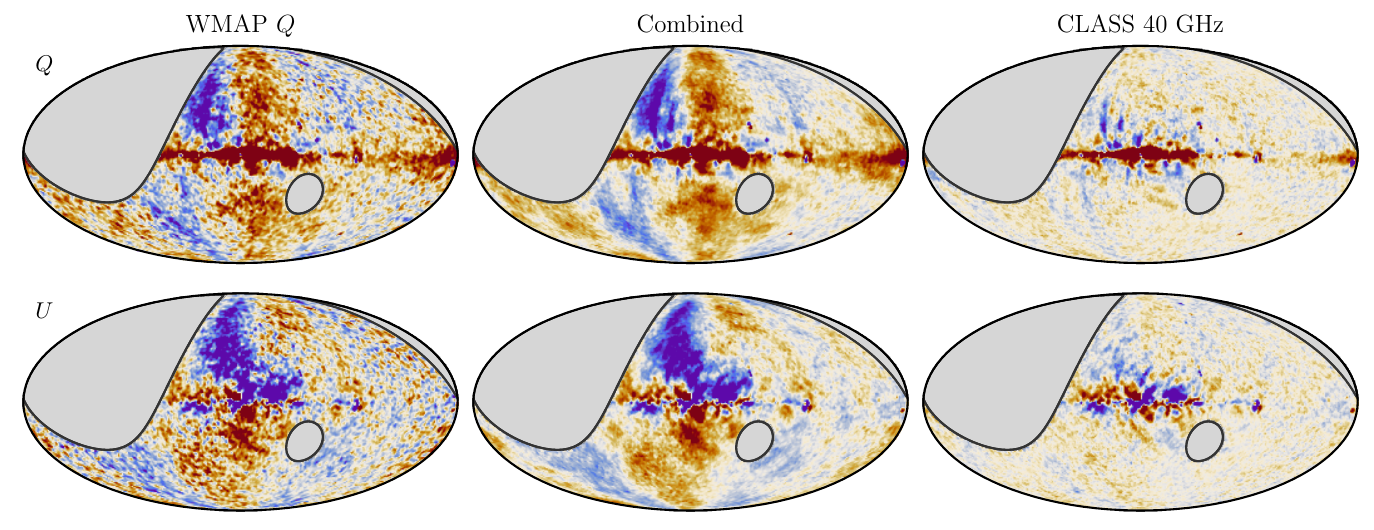}
    \caption{From left to right: the WMAP $Q$ band, combined and CLASS $40~\GHz$ maps.
    The top (bottom) panels show the Stokes $Q$ ($U$) in a linear color scale, and all panels share the same color range.
    The gray regions are beyond the CLASS survey boundary.
    Maps are converted to antenna temperature at $40~\GHz$ (Table \ref{tab:all_data}) and smoothed to FWHM=$2^\circ$ for visualization purposes.
    The combined maps take advantage of accurate information at the largest angular scales from \wq{} and high-sensitivity information at intermediate scales from C40 (improved by a factor of $2-3$ compared to \wq{} noise level).}
    \label{fig:multimaps}
\end{figure*}
In this section, we show the bandpass properties of the combined maps in Section \ref{ssec:bandpass} and present the maps and noise power spectra in Section \ref{ssec:maps}.

\subsection{Bandpass}\label{ssec:bandpass}
As the combination was done in the harmonic space and the weight is a function of both $\ell$ and $m$ (Figure \ref{fig:weights}), the bandpass of the combined maps depends on both $\ell$ and $m$, and can be computed following the same weighted average method as Equation \ref{eq:combine}.
In the left panel of Figure \ref{fig:bandpass} we show examples of the bandpass at several different $\ell$'s ($m$'s fixed at 15), and in the right panel we depict the band center of the combined maps as a function of $\ell$ for CMB (differential blackbody) and synchrotron (power law with $\beta_s=-3.1$) spectra.

The $\ell$ dependency of the combined maps band center mirrors the trend of the weights (Figure \ref{fig:weights}): it initially shifts rapidly toward the CLASS $40~\GHz$ band center at $\ell\sim15$ due to the significantly lower noise level of C40, and then gradually returns toward the W$Q$ band center as its angular resolution ($\mathrm{FWHM}\sim0.51^\circ$, corresponding to $\ell\sim350$) is higher than C40 ($\mathrm{FWHM}\sim 1.56^\circ$, corresponding to $\ell\sim115$) (Table \ref{tab:all_data}).
The $m$ dependency also follows the weights pattern (Figure \ref{fig:weights}), and the spread of the band center per $\ell$ is reflected by the darker shaded region in the right panel of Figure \ref{fig:bandpass}.
Taking into account the $0.5~\GHz$ uncertainty on the CLASS band center \citep{dahal22}, the total shift of the band center amounts to $2-3\GHz$, corresponding to a $5-7.5\%$ fractional shift.

\subsection{Maps and noise simulations}\label{ssec:maps}
The combined maps are shown in Figure \ref{fig:maps}; maps were smoothed to have an FWHM=$2^\circ$ Gaussian beam for visualization purposes.
The maps cover $\sim70\%$ of the sky, with boundary at declinations $\delta=-75^\circ$ and $\delta=29^\circ$.
The polarization intensity ($P$) and polarization angle ($\psi$) are visualized in the bottom panel, calculated as:
\begin{equation}
\begin{aligned}
    P&=\begin{cases}
        \sqrt{Q_1Q_2+U_1U_2} & \text{if $Q_1Q_2+U_1U_2\geq0$}\\
        0 & \text{else}
    \end{cases}\\
    \psi&=\frac12\mathrm{atan2}(U, Q).
\end{aligned}\label{eq:ppsi}
\end{equation}
To mitigate the noise bias, $P$ was computed from the two combined splits, with pixels having $Q_1Q_2+U_1U_2<0$ being zeroed out.
$\psi$ was directly computed from the combined maps.
The polarization intensity/angle map exhibits consistency with similar maps in previous studies \citep{planck2014-a12, planck2014-a31}.

\begin{figure*}[!t]
    \centering
    \includegraphics[width=\linewidth]{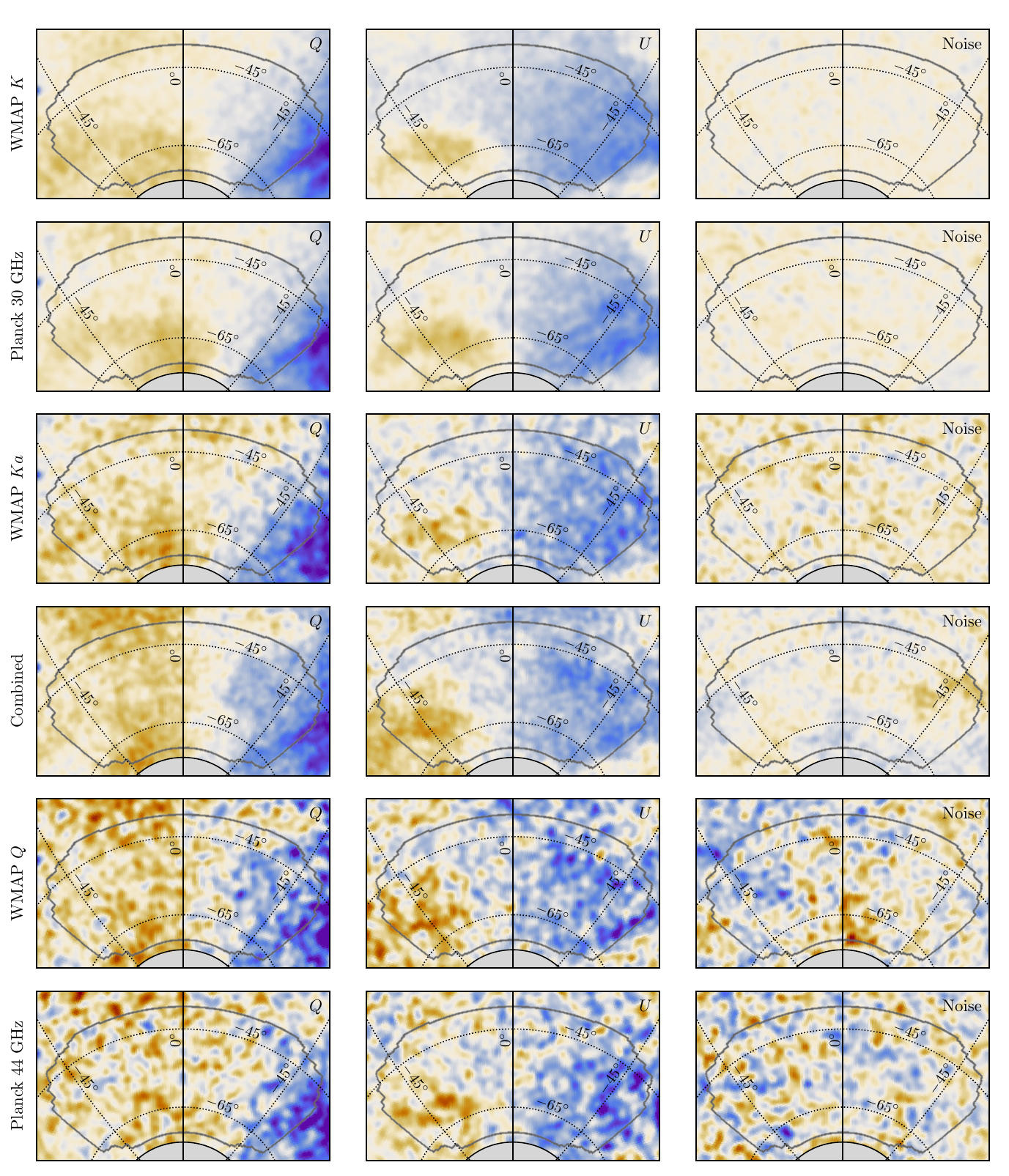}
    \caption{A comparison of data centered on BICEP3 region.
    The maps are presented in Lambert azimuthal equal-area projection, using the Celestial coordinate system.
    From top to bottom: the \wmap{} $K$, \planck{} 30 GHz, \wmap{} $Ka$, combined, \wmap{} $Q$ and \planck{} 44 GHz band data.
    Within each column, the maps display Stokes $Q$, Stokes $U$, and a single realization of Stokes $Q$ noise, respectively.
    The gray contours delineate the BICEP3 survey boundary, and the gray regions are beyond the CLASS survey boundary.
    All maps are converted to antenna temperature at 40 GHz (Table \ref{tab:all_data}) and smoothed to FWHM$=2^\circ$ for visualization purposes.
    Notably, the maps exhibit similar large-scale structures, predominantly influenced by polarized synchrotron radiation.
    The combined noise level is lower compared to those of the WMAP $Q$ and WMAP $Ka$ bands at intermediate scales when assuming $\beta_s=-3.1$.
    }
    \label{fig:multimaps_zoomin}
\end{figure*}

In Figure \ref{fig:multimaps} we compare the combined maps to the WMAP $Q$ and CLASS $40~\GHz$ band data. 
The maps have been converted to antenna temperature at $40~\GHz$ following a $\beta_s=-3.1$ power law.
Features at the largest scales ($\ell<15$, pixel size $>12^\circ$) in the combined maps mainly originate from \wq, while the map sensitivity level improves by a factor of $2-3$ at intermediate multipole ranges ($30<\ell<120$) owing to the lower noise level of C40.
Additionally, we compare the combined maps to more low-frequency band data focusing on the BICEP3 region\footnote{Mask available at \href{http://bicepkeck.org/bk18_2021_release.html}{http://bicepkeck.org/bk18\_2021\_release.html}. Instead of apodizing, we adopted a binary mask, including regions wherever the mask values are not \ttt{NaN}. We also excluded regions masked by the CLASS bright source mask \citep{eimer23}.} \citep{ade2022bicep} in Figure \ref{fig:multimaps_zoomin} (maps converted to antenna temperature at $40~\GHz$ similarly).
The Stokes $Q$/$U$ maps show similar large-scale structures as they are dominated by the polarized synchrotron radiation.
The combined noise level is lower than the WMAP $Q$ and $Ka$ band ones at intermediate scales when assuming $\beta_s=-3.1$.

\begin{figure}[t]
    \centering
    \includegraphics[width=\linewidth]{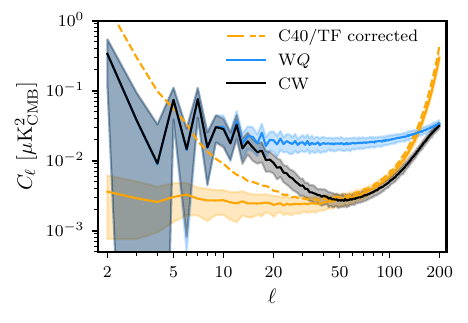}
    \caption{The $EE$ noise spectra for C40 (orange), \wq{} (blue), and the combined data (black). 
    Note that the `TF corrected' refers to the correction made by the power spectrum domain transfer matrix in \citet{Li23}, not the harmonic domain filtering matrix in this work.
    The noise spectra of the combined data generally track the lower of \wq{} or the TF corrected C40, but diverge at $\ell\in(7,40)$ mainly due to an additional high-pass filter (Equation \ref{eq:weights}).
    The $BB$ spectra are qualitatively the same.}
    \label{fig:noise_aps}
\end{figure}

We compare the $EE$ power spectra of the combined noise simulations to that for CLASS $40~\GHz$ and WMAP $Q$ band in Figure \ref{fig:noise_aps}.
In general, the noise spectra of the combined data track the lower of WQ or the TF corrected C40,\footnote{The `TF corrected' refers to the correction made by the power spectrum domain transfer matrix in \citet{Li23} not the harmonic domain filtering matrix in this work.} but diverge slightly at $\ell\in(7, 40)$ mainly due to an additional high-pass filter applied on the C40 weights (Equation \ref{eq:weights}).
The spectra were computed using \ttt{PolSpice} \citep{polspice}, and the $BB$ spectra are qualitatively the same.

\section{Pixel-based polarized synchrotron spectral indices analysis}\label{sec:beta_s}
The synchrotron radiation is an important source of polarized diffuse Galactic emission at frequencies lower than $\sim 70~\GHz$ \citep{krachmalnicoff2018spass, weiland2022polarized}, and dominates over the polarized CMB signals on angular scales larger than degree-level. 
Understanding the synchrotron radiation is crucial for accurately performing component separation \citep{planck2016-l04,hoz2023quijote,svalheim2023beyondplanck,watts2023cosmoglobe}.
Typically, its frequency spectrum can be approximated by a power law ($T \propto \nu^{\beta_s}$) over a certain range of frequencies, where $\beta_s$ is the polarized synchrotron spectral index. 

In this section, we present a spatial variation analysis on $\beta_s$ with the assistance of the WMAP $K$ and $Ka$ band data.
We introduce the method to fit for $\beta_s$ in Section \ref{ssec:tt_method}, summarize the fitting pipeline in Section \ref{ssec:beta_s_process}, and discuss the results in Section \ref{ssec:beta_s_result}.

\subsection{Method}\label{ssec:tt_method}
We employed the commonly used $T-T$ plot method for fitting the $\beta_s$ values \citep{fuskeland2014spatial, fuskeland2021constraints, weiland2022polarized, eimer23}, by fitting a zero-intercept line to a scatter plot where the $x$- and $y$-axes are the Stokes $Q$ and $U$ from different frequency channels.
In the presence of uncertainty on both axes, an unbiased estimation of the best-fit slope can be obtained using the total least-squares fitting as:
\begin{equation}
    \begin{aligned}
    \hat k&=\underset{k}{\mathrm{argmin}}\left(-2\log\mathcal L\right)\\
    &=\underset{k}{\mathrm{argmin}}\left[\pmb{d}(k)^T\ms C^{-1}\pmb{d}(k)\right],\label{eq:beta_tls}
    \end{aligned}
\end{equation}
where $-2\log\mathcal L\equiv \pmb{d}(k)^T\ms C^{-1}\pmb{d}(k)$ is the log-likelihood, $\pmb d(k)\equiv\pmb d_y-k\pmb d_x$, and $\ms C\equiv k^2\ms C_x+\ms C_y$ is the covariance matrix.
$\pmb d_{x, y}$ are the Stokes $Q$ and $U$ data from different experiments, and $\ms C_{x,y}$ are the $Q$/$U$ covariance matrices obtained from CMB $+$ noise simulations as described in Section \ref{sec:data}.
The subscript $_x$ refers to \wk\ data, and $_y$ refers to either \wka\ or CW data.
The fitted slope can be converted to the $\beta_s$ value as:
\begin{equation}
    \hat\beta_s=\frac{\log{\hat k}}{\log(\nu_y/\nu_x)},\label{eq:betas}
\end{equation}
where the referenced frequencies $\nu_{x,y}$ can be found in Table~\ref{tab:all_data}.
We used the original and reobserved \planck{} PR4 353~GHz maps as the template for polarized thermal dust radiation.
They were subtracted (in thermodynamic temperature units) from corresponding bands with scaling factors in Table \ref{tab:all_data} applied.

The uncertainty on the $\hat\beta_s$ was transformed from the uncertainty on $\hat k$:
\begin{equation}
\begin{aligned}
    \Delta k^\mathcal{L}&=\left({\pmb d}_x^T\hat{\ms C}^{-1}\pmb d_x\right)^{-1/2},\\
    \Delta\beta_s^\mathcal{L}&=\frac{\Delta k^\mathcal{L}}{\hat k\log(\nu_y/\nu_x)},
\end{aligned}\label{eq:dbeta_s}
\end{equation}
where $\hat{\ms C}=\hat k^2\ms C_x+\ms C_y$, and $\Delta k^\mathcal{L}$ is the square root of the inverse Fisher information of the log-likelihood where $\ms C$ was fixed to be $\hat{\ms C}$ in the derivation.
We use ensemble simulations to test the accuracy of $\Delta\beta_s^\mathcal{L}$, and found that they need to be corrected as $1.1\times\Delta\beta_s^\mathcal{L}$ for the \wk--\wka\ fitting and $\Delta\beta_s^\mathcal{L}+0.05$ for the \wk--CW fitting (Appendix \ref{sec:beta_s_verify}).
For pixels with probability-to-exceed (PTE) values lower than 0.01, we replace the corrected $\Delta\beta_s^\mathcal L$ with the bootstrapped uncertainty if the latter is larger.
We fit $\hat k$ for 5000 different data resamplings using only the diagonal components of the covariance matrix (resampled in the same way), and the standard deviation of the spectral indices inferred from different samples is treated as the bootstrapped uncertainty.

The $\beta_s$ values are fitted at \ttt{HEALPix} resolution $\nside=8$ (pixel size $\sim7.3^\circ$) as was adopted in \citet{weiland2022polarized} and \citet{eimer23}.
All maps and simulations are smoothed to a common FWHM=$2^\circ$ Gaussian beam size and downgraded to $\nside=32$ for this analysis.
Since the information at $\ell,m>383$ was significantly suppressed in this analysis, we used the combined maps that were created at $\nside=128$ instead of $\nside=256$, i.e., without appending the weighted averaged spherical harmonics coefficients with the coefficients of the \wq{} data beyond $\ell,m=383$.
We refer the reader to \citet{eimer23} for further details, but we note that the mask adopted here is slightly different: at $\nside=128$ we used a declination range from $-74^\circ$ to $28^\circ$, and combined it with the smoothed bright source mask (from C40 beam profile to FWHM=$2^\circ$ Gaussian), then setting all values smaller than 0.9 to be zero, otherwise unity.
The mask is then processed in the same way as in \citet{eimer23}.
This slightly more conservative mask ensures that the residual from the combined method (Appendix \ref{ssec:sync_verify}) does not significantly bias the $\beta_s$ values.

\subsection{Fitting process}\label{ssec:beta_s_process}
\begin{figure*}
    \centering
    \includegraphics[width=\linewidth]{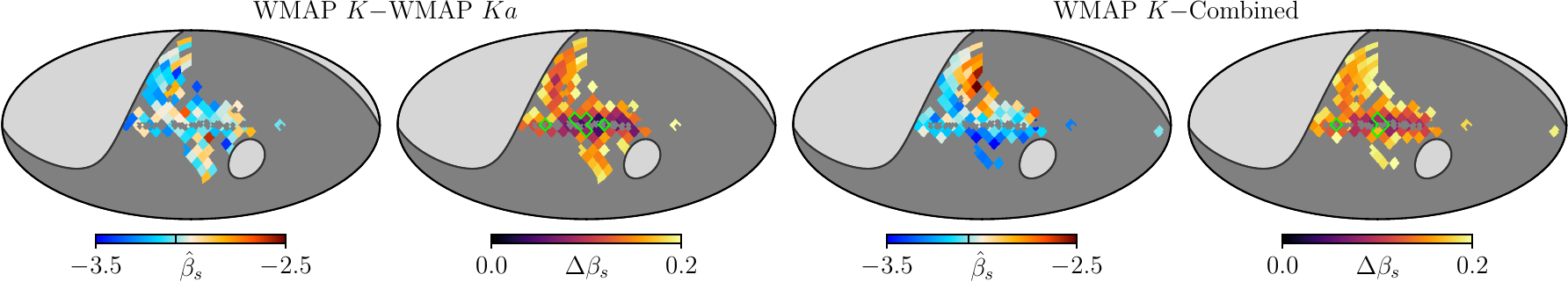}
    \caption{The $\beta_s$ and $\Delta\beta_s$ maps from \wk$-$\wka{} (left two panels) and \wk$-$CW (right two panels) with resolution $\nside=8$ (pixel size $\sim 7.3^\circ$).
    Regions with $\Delta\beta_s>0.2$ and those masked by the declination limit (light gray) and point-source mask are excluded.
    In the second and fourth panels from the left, pixels with $\mathrm{PTE}<0.01$ are highlighted by green borders.
    The mean and 16/84th percentiles are $-3.08_{-0.14}^{+0.14}$ for $\beta_s^{\mr WK-\mr WKa}$ and $-3.08_{-0.20}^{+0.20}$ for $\beta_s^{\mr WK-\mr{CW}}$.
    The black vertical lines in the $\beta_s$ map color bars mark the mean value across each map.
    In the $\Delta\beta_s$ maps, pixels highlighted by green borders have their values determined by bootstrapping.
    The $\beta_s^{\mr WK-\mr{CW}}$ map prefers spatial variation on the spectral index (PTE for a uniform $\beta_s$ hypothesis smaller than 0.001).
    }
    \label{fig:bs}
\end{figure*}
Acquiring spatial variation information for the SED when the bandpass differs at each $\ell$ and $m$ is not trivial.
Instead of implementing complex convolution processes to convert the combined data to antenna temperature, we first converted \wq\ and C40 separately and then combined them.
Here is the process we used to obtain $\beta_s$ maps:
\begin{enumerate}[itemsep=0pt]
    \item Begin by assuming $\beta_s=-3.1$ all over the sky.
    \item Convert the thermal dust subtracted \wk\ band data to antenna temperature at reference frequency ($22.8~\GHz$) following the $\beta_s$ map.
    \item Convert the thermal dust subtracted \wq\ and C40 band data to antenna temperature at $40~\GHz$ following the $\beta_s$ map.
    \item Combine the converted C40 and \wq{} band data following Equation \ref{eq:combine}. The combination was done with $\nside=128$ maps directly because the information at $\ell>383$ is strongly suppressed in this analysis.
    \item Fit for a $\beta_s$ map using the method described in Section \ref{ssec:tt_method}.
    \item Iterate steps 2--5 until the difference in $\beta_s$ per pixel between the current and last iteration is smaller than 0.1\% (typically reached within 5 iterations).
\end{enumerate}

The harmonic domain noise (Section \ref{ssec:noise}) and the CMB + noise simulations were converted to antenna temperature assuming a $\beta_s=-3.1$ power law and remained fixed throughout the fitting process.
This approach is computationally efficient and only induces negligible bias in the $\hat\beta_s$ and $\Delta\beta_s$.

\subsection{$\beta_s$ fitting results}\label{ssec:beta_s_result}
\begin{figure}
    \centering
    \includegraphics[width=\linewidth]{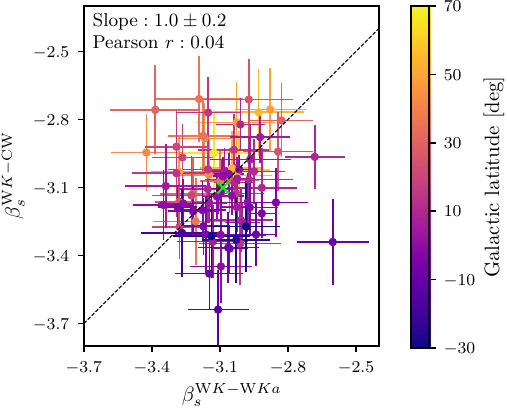}
    \caption{The scatter of $\beta_s$ fitted from two pairs of data, using pixels available in both pairs.
    The green cross marks the mean values of the two $\beta_s$ maps.
    The black dashed line is plotted for reference ($y=x$).
    By fitting a line with zero-intercept, we found its slope ($1.0\pm0.2$) is consistent with unity, showing no preference for frequency dependence on $\beta_s$.
    The Pearson correlation coefficient is $0.04$, showing no significant correlation between two $\beta_s$ maps.
    The data points are color-coded by the Galactic latitudes, which seems to correlate with $\beta_s^{\mr WK-\mr{CW}}$ values (this interpretation requires caution due to the large uncertainty).}
    \label{fig:beta_comparison}
\end{figure}

Figure \ref{fig:bs} shows the $\beta_s$ maps and their uncertainty $\Delta\beta_s$ maps for two pairs of data: \wk{}--\wka\ and \wk{}--CW.
Only pixels with $\Delta\beta_s<0.2$ are shown.
The green-bordered pixels in the uncertainty maps had their values determined by bootstrapping (Section \ref{ssec:tt_method}).
The results for \wk{}--\wka{} are mostly identical to those in \citet{weiland2022polarized} except for regions in the Galactic plane due to a different mask.

The mean and 16/84th percentile value for $\beta_s^{\mr WK-\mr WKa}$ is $-3.08_{-0.14}^{+0.14}$ and for $\beta_s^{\mr WK-\mr{CW}}$ is $-3.08_{-0.20}^{+0.20}$, which are consistent with previous studies that computed $\beta_s$ along the plane \citep{wmap3kogut07, ruud2015imaging, weiland2022polarized, eimer23}.
The PTE for all $\beta_s$ with $\Delta\beta_s<0.2$ to be consistent with their mean (where $\chi^2_\mr{uniform}\equiv\sum[(\beta_s-\bar{\beta}_s)^2/\Delta\beta_s^2]$ and $\bar{\beta}_s$ is the mean across the map) is 0.01 for \wk{}--\wka{} and smaller than 0.001 for \wk{}--CW.
(The $\mr WK-\mr WKa$ maps have 91 $\nside=8$ pixels with $\Delta\beta_s<0.2$, and $\mr WK-\mr{CW}$ maps have 85.)
The PTE for $\beta_s^{\mr WK-\mr WKa}$ to be consistent with $\beta_s^{\mr WK-\mr{CW}}$ is 0.07, of which the $\chi^2$ is defined to be:
\begin{equation}
    \chi^2=\sum\frac{\left(\beta_s^{\mr WK-\mr WKa}-\beta_s^{\mr WK-\mr{CW}}\right)^2}{\left(\Delta\beta_s^{\mr WK-\mr WKa}\right)^2+\left(\Delta\beta_s^{\mr WK-\mr{CW}}\right)^2},
    \label{eq:interchi2}
\end{equation}
summed over all the common pixels (80 in total).
The slightly low PTE value indicates that the $\beta_s$ fitted from two data pairs are slightly in tension, and the \wk{}--CW pair prefers spatial variation on $\beta_s$.

In Figure \ref{fig:beta_comparison} we compare $\beta_s^{\mr WK-\mr WKa}$ and $\beta_s^{\mathrm{W}K\mathrm{-CW}}$ using data from pixels that are available in both $\beta_s$ maps, color-coded by Galactic latitudes.
The apparent correlation observed between $\beta_s^{\mr WK-\mr{CW}}$ and the Galactic latitudes (noticeable in Figure \ref{fig:bs}) should be approached with caution due to the large uncertainty.
The Pearson correlation coefficient $r$ is $0.04$, indicating no significant correlation between the $\beta_s$ from two pairs of data.
Additionally, we computed the Pearson correlation coefficient for pixels with $\Delta\beta_s<0.15$ (36 common pixels), yielding a value of $0.26$. 
This suggests that the low correlation coefficient in Figure \ref{fig:beta_comparison} is likely due to the instrumental noise.
We fitted a line with zero-intercept using the same method as in Equation \ref{eq:beta_tls}, where the covariance matrix is diagonal in this case. 
The best-fit slope is $1.0\pm0.2$, consistent with unity, showing no evidence for potential frequency dependence on $\beta_s$ when fitted in the current form and using \wk{}, \wka{}, and CW data. 
~\\

In short, the $\beta_s$ fitted from \wk{}--CW is consistent with several existing studies \citep{kogut2007, ruud2015imaging, weiland2022polarized, eimer23}.
Due to the sensitivity improvement at intermediate scales introduced by the CLASS $40~\GHz$-band data, the $\Delta\beta_s$ fitted with the combined $40~\GHz$ map is comparable to that with the WMAP $Ka$ band map.
However, the uncertainty in the absolute calibration at CLASS $40~\GHz$ data is not negligible for this $\beta_s$ analysis because the frequency leverage arm is short.
It has been shown in \citet{eimer23} that a shift in calibration by $\pm 5\%$ is equivalent to shifting the $\beta_s$ measurement in all $N_{\rm side}=8$ pixels by a common $\pm 0.1$.
As no significant difference was noticed between $\beta_s^{\mr WK\mr{-CW}}$ and $\beta_s^{\mr WK-\mr WKa}$, we refer readers to \citet{weiland2022polarized} for a detailed analysis of the implications for foreground removal.


\section{Conclusion}\label{sec:conclusion}
In this work, we made sensitivity-improved polarization maps at $40~\GHz$ by combining the CLASS $40~\GHz$ band and WMAP $Q$ band-data.
The combined maps were made at $\nside=256$ with a smoothing scale of FWHM=$1^\circ$, by doing an inverse noise variance weighted average in the harmonic space at $\ell, m\leq383$ and filling the $\ell,m>383$ with WMAP data (Section \ref{sec:method}).
We used a harmonic space filtering matrix to replicate the filtering inherent in the CLASS $40~\GHz$ data pipeline.
An additional high-pass filter was applied to the C40 weights to ensure the bias from the combination method was negligible.
We validated the combination method by comparing the binned data cross spectra (Appendix \ref{sec:dataspectra}), finding that the spectra of the combined maps follow a similar trend to other low-frequency data.
The CMB and realistic sky simulation checks in Appendix \ref{sec:verification} validate that the combination method results in negligible bias.

We performed a pixel space $T-T$ plot polarized synchrotron spectral index analysis with the combined maps. 
We found that the $\beta_s$ map (\ttt{HEALPix} resolution $\nside=8$, pixel size $\sim7.3^\circ$) estimated from WMAP $K$ and the combined data ($\beta_s=-3.08^{+0.20}_{-0.20}$) shows a stronger preference for spatial variation than that from  WMAP $K$ and WMAP $Ka$ ($\beta_s=-3.08^{+0.14}_{-0.14}$), with the former showing correlation with the Galactic latitudes (should be approached with caution due to the large $\Delta\beta_s$).
No frequency relation was found from the two $\beta_s$ maps.
In the future, it is worth including \planck{} LFI data and data from other experiments such as S-PASS \citep{carretti2019sband}, C-BASS \citep{jones2018cband}, and QUIJOTE \citep{rubino2023quijote} to extend the leverage arm, which holds promise in revealing the potential frequency dependence of $\beta_s$ (a steepening of $\beta_s$ from the intensity data was noticed in \citealt{kogus2012synchrotron}).
The combined data also hold the potential to broaden our knowledge of any polarized AME.

The products of this work, including the combined maps, the combined noise simulations, the harmonic domain filtering matrix, and the $\beta_s$ maps, are publicly available on \ttt{LAMBDA}.\footnote{\href{https://lambda.gsfc.nasa.gov/product/class/class_prod_table.html}{https://lambda.gsfc.nasa.gov/product/class/class\_prod\_table.html}}
Other products can be made available upon request.

\section{Acknowledgments}
We acknowledge primary funding support for CLASS from the National Science Foundation Division of Astronomical Sciences under Grant Numbers 0959349, 1429236, 1636634, 1654494, 2034400, and 2109311. We thank Johns Hopkins University President R. Daniels and the Deans of the Kreiger School of Arts and Sciences for their steadfast support of CLASS. We further acknowledge the very generous support of Jim and Heather Murren (JHU A\&S '88), Matthew Polk (JHU A\&S Physics BS '71), David Nicholson, and Michael Bloomberg (JHU Engineering '64). The CLASS project employs detector technology developed in collaboration between JHU and Goddard Space Flight Center under several previous and ongoing NASA grants. Detector development work at JHU was funded by NASA cooperative agreement 80NSSC19M0005. CLASS is located in the Parque Astron\'omico Atacama in northern Chile under the auspices of the Agencia Nacional de Investigaci\'on y Desarrollo (ANID).
We acknowledge scientific and engineering contributions from 
Max Abitbol, 
Aamir Ali,
Fletcher Boone,
Michael~K. Brewer,
Sarah~Marie Bruno,
David Carcamo, 
Carol Chan, 
Manwei Chan, 
Mauricio D\'iaz,
Kevin~L. Denis,
Francisco Espinoza, 
Benjam\'in Fern\'andez, 
Pedro Flux\'a Rojas,
Joey Golec, 
Dominik Gothe, 
Ted Grunberg, 
Mark Halpern, 
Saianeesh Haridas, 
Kyle Helson, 
Gene Hilton, 
Connor Henley, 
Johannes Hubmayr,
John Karakla,
Lindsay Lowry, 
Jeffrey~John McMahon, 
Nick Mehrle, 
Nathan J.~Miller,
Carolina~Morales Perez, 
Carolina N\'{u}\~{n}ez,
Keisuke Osumi,
Ivan~L. Padilla, 
Gonzalo Palma, 
Lucas Parker, 
Sasha Novack, 
Bastian Pradenas, Isu Ravi, 
Carl~D. Reintsema, 
Gary Rhoades, Daniel Swartz, Bingjie Wang, Qinan Wang, 
Tiffany Wei, 
Zi\'ang Yan,
Lingzhen Zeng, 
and Zhuo Zhang. 
For essential logistical support, we thank Jill Hanson, William Deysher, Joseph Zolenas, LaVera Jackson, Miguel Angel D\'iaz, Mar\'ia Jos\'e Amaral, and Chantal Boisvert. 
We acknowledge productive collaboration with the JHU Physical Sciences Machine Shop team.
R. S. thanks Sihao Cheng for the helpful discussion.
R. D\"unner thanks ANID for grant BASAL CATA FB210003.
Z. X. is supported by the Gordon and Betty Moore Foundation through grant GBMF5215 to the Massachusetts Institute of Technology.

\software{
numpy \citep{numpy20}, 
scipy \citep{scipy}, 
matplotlib \citep{matplotlib},
astropy \citep{astropy}, 
HEALPix \citep{healpix},
fastcc \citep{fastcc},
pysm \citep{pysm},
PolSpice \citep{polspice},
}

\setcounter{figure}{0}
\renewcommand{\thefigure}{A\arabic{figure}}
\appendix
\placefigure{fig:aps}
\section{Data power spectra}\label{sec:dataspectra}
\begin{figure*}[t!]
    \centering
    \includegraphics[width=\linewidth]{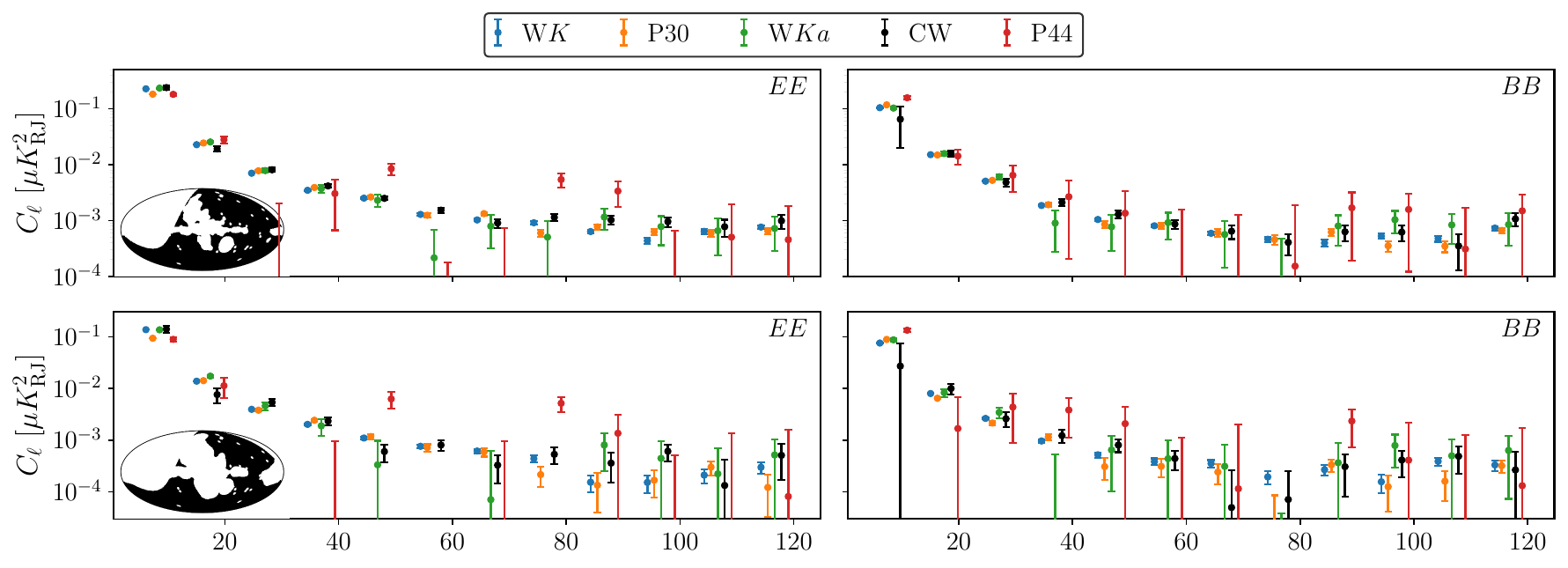}
    \caption{The binned cross power spectra from \wk{} (blue), P30 (orange), \wka{} (green), CW (black), and P44 (red), with multipole range from $\ell=2$ to $122$, binned per $\Delta\ell=10$, and weighted by $\ell(\ell+1)$.
    All spectra have been converted to antenna temperature and scaled to $40~\GHz$ following a $\beta_s=-3.1$ power-law frequency spectrum.
    The \planck{} 2018 best-fit theory spectra have been subtracted.
    \textit{Top left}: $EE$ spectra with \ttt{s5} mask applied.
    \textit{Top right}: $BB$ spectra with \ttt{s5} mask applied.
    \textit{Bottom left}: $EE$ spectra with \ttt{s9} mask applied.
    \textit{Bottom right}: $BB$ spectra with \ttt{s9} mask applied.
    Masks are shown on the bottom left of the left panels.
    Only the instrumental noise is included in the error bar size, not the signal sample variance.
    }
    \label{fig:aps}
\end{figure*}
We show the binned cross power spectra of WMAP $K$, $Ka$, Planck $30~\GHz$, $44~\GHz$-band maps and the combined maps in Figure \ref{fig:aps}.
The power spectra were estimated by \ttt{PolSpice}, between the coadded single-year splits for WMAP and the half-ring splits for \planck{} (Section \ref{sec:data}).
For the combined data we used the two splits described in Section \ref{ssec:summary}.
The multipole range was from $\ell=2$ to $\ell=122$ (beyond $\ell=122$ the combined data is dominated by the noise) and the power spectra were binned per $\Delta\ell=10$, weighted by $\ell(\ell+1)$.
The \planck{} 2018 best-fit theory spectra for the CMB have been subtracted from the binned cross spectra.
The power spectra are scaled to $40~\GHz$ (assuming $\beta_s=-3.1$) for comparison.
The error bars are the ensemble standard deviation of 200 noise cross spectra, and they reflect only the instrumental noise, not the signal sample variance.

We compared the spectra for two cases with different masks (\ttt{s5} and \ttt{s9}) applied, and they are qualitatively the same.
At the largest angular scales ($\ell<20$) the CW data are dominated by the \wq{}, so the error bar size in the first two bins is larger compared to those from \wka.
The CW error bar size at $\ell>20$ is generally smaller than that from the \wka, implying that the CW data has a higher signal-to-noise ratio at those scales.
Note that this comparison aims to validate the combination method---the CW spectra follow a similar shape as other low-frequency band data do.
This is not to claim that the power spectra from different bands should all be consistent.

\setcounter{figure}{0}
\renewcommand{\thefigure}{B\arabic{figure}}
\begin{figure*}
    \centering
    \includegraphics[width=\linewidth]{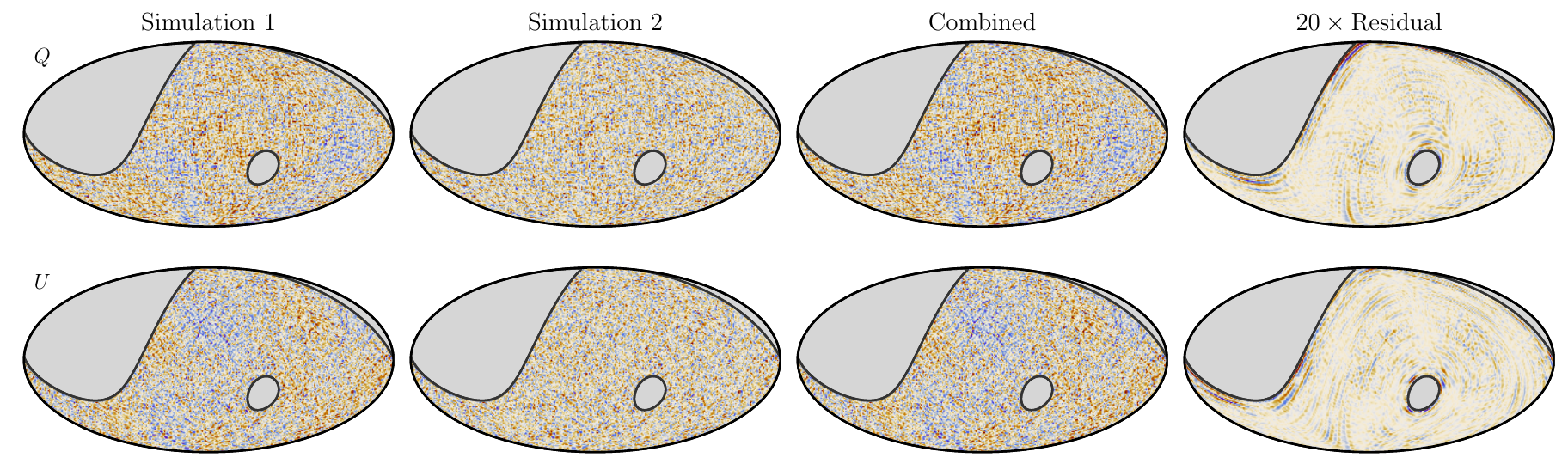}
    \caption{The CMB simulation verification in the map domain (a typical one of 200 realizations). 
    All maps share the same color range.
    The gray regions are beyond the CLASS survey boundary.
    From left to right: maps representing \wq{}-like signal, maps representing C40-like signal, the combined maps using weights applied to the real data, and 20 times the residual between the combined and the Simulation 1.
    The signal maps were created at the same monochromatic frequency.
    The residual is negligible compared to the signal.}
    \label{fig:cmbr_verify_map}
\end{figure*}
\section{Combination method validation with simulations}\label{sec:verification}
In this section, we describe the simulation checks conducted to validate our combination method.
\subsection{CMB only simulations}\label{ssec:cmbr_verify}
We began with CMB-only simulations.
We followed the method described in Section \ref{ssec:sims} when generating the CMB simulations.
These simulations were treated as the observation made by the WMAP $Q$ band.
For the observation made by CLASS $40~\GHz$, we used the reobserved CMB simulations done with CLASS 40 GHz data pipeline \citep{Li23}.
In this analysis, no bandpass differences were assumed.
Our primary objective was to assess the residual effects resulting from the combination method, rather than investigating imperfections in our understanding of the bandpass.

\begin{figure}
    \centering
    \includegraphics[width=\linewidth]{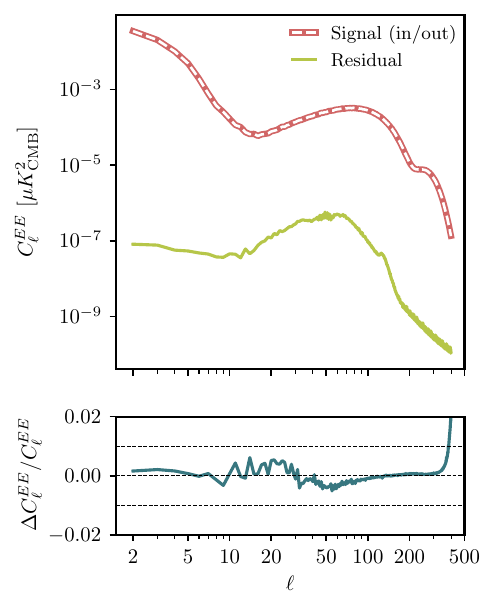}
    \caption{The $EE$ power spectra (CLASS survey boundary mask applied) of the CMB simulations.
    \textit{Top}: the signal (thick red line for the input, white dashed line for the combined signal) and residual (green), obtained as the ensemble average over 200 simulations.
    The residual power is negligible compared to the signal.
    \textit{Bottom}: the fractional difference between input and combined spectra, mostly within 1\% at $\ell<380$.
    The signal at $\ell\geq380$ is significantly suppressed by the beam size (FWHM=$1^\circ$, $\ell\sim180$).}
    \label{fig:cmbr_verify_aps}
\end{figure}

By combining the two different CMB simulations in the same way we did with the data, the resulting maps and power spectra are shown in Figure \ref{fig:cmbr_verify_map} and \ref{fig:cmbr_verify_aps}.
The residual levels in both the maps and the power spectra are negligible compared to the signal.

\subsection{Realistic simulations}\label{ssec:sync_verify}
We then performed validation checks with more realistic simulations.
We included 4 components in our signal simulation:
\begin{itemize}[itemsep=-2pt]
    \item Polarized synchrotron: we used the \pysm{} \ttt{s1} model (not to be confused with the CLASS \ttt{s1} mask) at 40~GHz, smoothed with an FWHM=$1^\circ$ Gaussian beam, in $\mu K_\mathrm{CMB}$.
    \item Polarized thermal dust: we used the \pysm{} \ttt{d1} model (not to be confused with the CLASS \ttt{d1} mask) at 40~GHz, smoothed with an FWHM=$1^\circ$ Gaussian beam, in $\mu K_\mathrm{CMB}$.
    \item CMB simulations: the same as in Appendix \ref{ssec:cmbr_verify}.
    \item Bright sources: we computed the bright source amplitudes (Stokes $Q$ and $U$) by aggregating all pixel values within each bright source outlined by the CLASS bright source mask. The point source maps were generated by assigning the total amplitudes to pixels closest to the center of the corresponding bright source mask holes. Finally, the maps were smoothed using an FWHM=$1^\circ$ Gaussian beam.
\end{itemize}
In this analysis, no bandpass differences were assumed, and no extra calibration was applied to the bright sources map.
We note that the bright sources map we made is neither complete nor fully accurate, but it represents the impact of the brightest sources on the sky.
We highlight again that our primary objective was to assess the residual effects resulting from the combination method, rather than investigating imperfections in our understanding of the bandpass, or uncertainty on the morphology of the different components.
We treat the signal simulation as the observation made by WMAP $Q$ band, and the reobserved \citep{Li23} signal map as the observation made by CLASS $40~\GHz$ band.

\begin{figure*}
    \centering
    \includegraphics[width=\linewidth]{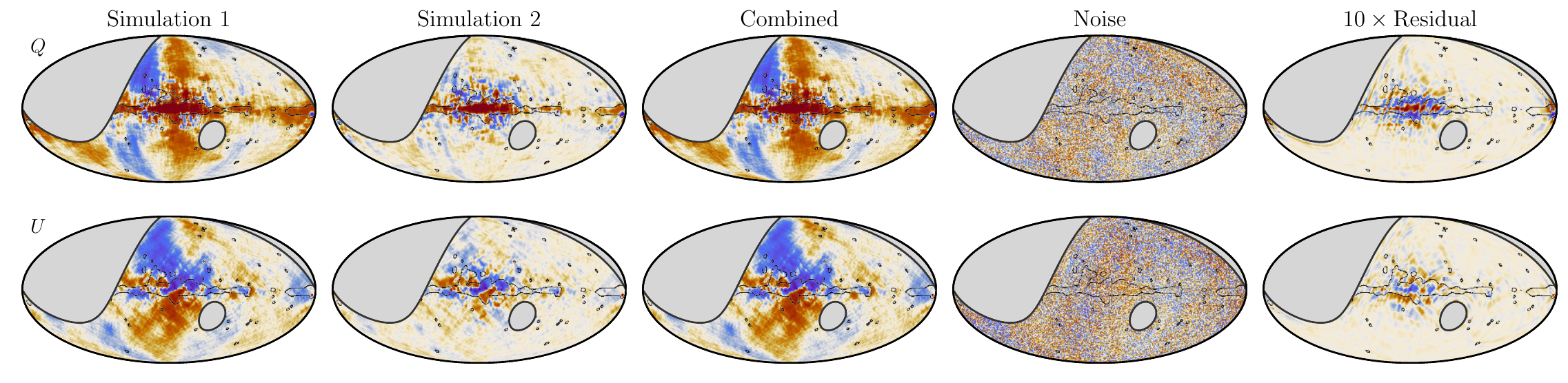}
    \caption{The realistic simulation verification in the map domain. 
    All maps share the same color range.
    The gray regions are beyond the CLASS survey boundary.
    From left to right: maps representing \wq{}-like signal, maps representing C40-like signal, the combined maps using weights applied to the real data, combined noise simulations, and 10 times the residual between the combined and the Simulation 1.
    The signal maps were created at the same monochromatic frequency.
    The residual in most regions of the sky is negligible compared to the signal.
    In the low Galactic latitudes, the residual is relatively higher due to the bright signal from the Galactic plane.
    The CLASS \ttt{s0+d0} mask (black borders, \citealt{eimer23}) captures most of the high residual regions, and the ratio between the standard deviation of pixel values in residual and Simulation 1 enclosed by the CLASS \ttt{s0+d0} mask is 3\%.
    }
    \label{fig:sync_verify_map}
\end{figure*}
\begin{figure*}
    \centering
    \includegraphics[width=\linewidth]{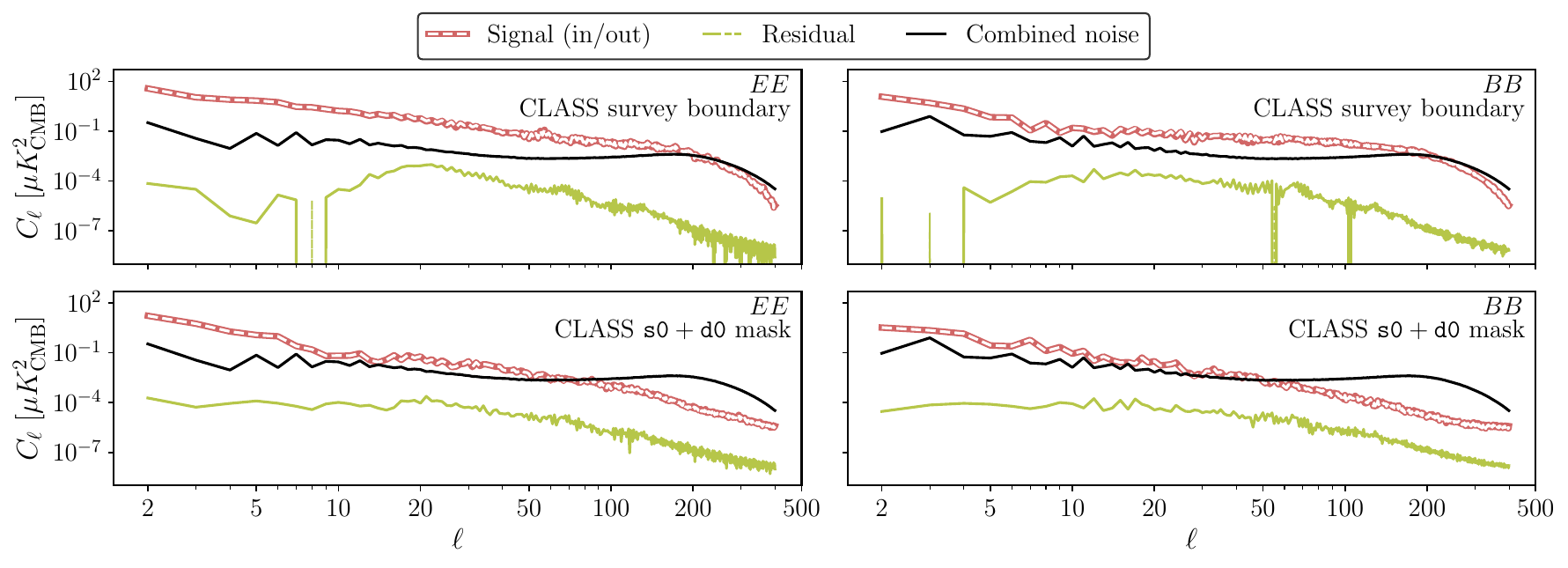}
    \caption{The power spectra of the realistic simulations: signal (thick red lines for the input, white dashed lines for the combined signal), residual (green, dashed parts are where the power spectra are negative), and combined noise (black, ensemble average of 200 simulations).
    \textit{Top}: with CLASS survey boundary mask applied.
    \textit{Bottom}: with CLASS \ttt{s0+d0} mask applied.
    In all the panels, the residual power is two orders of magnitude smaller than either the signal or the noise level.}
    \label{fig:sync_verify_aps}
\end{figure*}

By combining the two different signal simulations in the same way as we did with the data, the resulting maps and power spectra are shown in Figure \ref{fig:sync_verify_map} and \ref{fig:sync_verify_aps}.
We also include one typical realization of the CW noise simulation for comparison.
The residual is negligible compared to the signal in most regions of the sky; however, in the low Galactic latitudes, the residual is relatively higher due to the bright signal from the Galactic plane, and we found that the combination of the CLASS \ttt{s0} and \ttt{d0} (\ttt{s0+d0}) mask (\citet{eimer23}) captures most of the high residual regions.
The ratio between the standard deviation of pixel values in residual and Simulation 1 enclosed by the CLASS \ttt{s0+d0} mask is 3\%.
In general, the residual levels in the map space are negligible compared to the signal, and residual power spectra are two orders of magnitudes smaller than either the noise or the signal.

\section{$\beta_s$ fitting validation}\label{sec:beta_s_verify}
In this section we validate the modification we made to $\Delta\beta_s^\mathcal L$ (Equation \ref{eq:dbeta_s}).

We used the \pysm{} \ttt{s1} model at $22.8~\GHz$ as the polarized synchrotron signal template (Rayleigh-Jeans temperature unit) and scaled it to $33~\GHz$ and $40.7~\GHz$ using the \ttt{s1} $\beta_s$ model.
These simulations were converted to the thermodynamic temperature unit, and they represent the synchrotron templates for WMAP $K$, $Ka$, and $Q$ bands.
For CLASS $40~\GHz$ band, we scaled the $22.8~\GHz$ template to $38~\GHz$ following the same \ttt{s1} model, converted it to the thermodynamic temperature unit, and treated the reobserved simulation as the template.
We generated an ensemble of 200 simulations by combining the signal amplitudes with the 200 CMB $+$ noise simulations for each of the bands; below and in Figure \ref{fig:beta_s_verify}, we refer to these simulations as the ``ensemble simulations'' and the spectral index derived therefrom as $\beta_s^{ens}$.
(For C40 we used the reobserved CMB simulations.)
We fitted for $\beta_s$ from signal-only simulations and the ensemble simulations following the same procedure as was done for the data.

In Figure \ref{fig:beta_s_verify} we compare the corrected $\Delta\beta_s^\mathcal L$ to the ensemble scattering.
The $\Delta\beta_s^\mathcal L$ were calculated using Equation \ref{eq:dbeta_s} with $\hat k$ fitted from signal-only simulations.
The ensemble scattering for the \wk{}--\wka{} pair was obtained as the standard deviation of the $\beta_s$ from 200 simulations.
For the \wk{}--CW pair, in addition to the standard deviation of the $\beta_s$ from 200 simulations, we also include the absolute difference between $\beta_s$ fitted from the signal-only simulations and the input $\beta_s$ model.
This reflects the bias caused by the residual resulted from the combined method (Figure \ref{fig:sync_verify_map}).

\setcounter{figure}{0}
\renewcommand{\thefigure}{C\arabic{figure}}
\begin{figure}
    \centering
    \includegraphics[width=\linewidth]{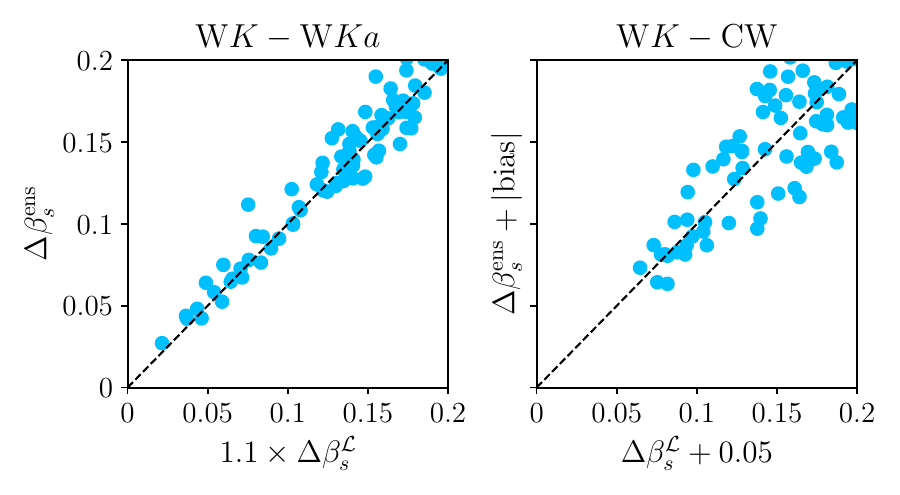}
    \caption{The scatter between corrected $\Delta\beta_s^\mathcal L$ and $\Delta\beta_s^\mathrm{ens}$. 
    \textit{Left}: \wk{}--\wka\ fitting. \textit{Right}: \wk{}--CW fitting.
    The black dashed lines are $y=x$ for reference.
    The data points gently scatter around the $y=x$ line, implying that the corrected $\Delta\beta_s^\mathcal L$'s capture the ensemble scattering caused by the noise.}
    \label{fig:beta_s_verify}
\end{figure}

\bibliography{main.bib,class_pub.bib,Planck_bib.bib,cmb.bib,software.bib}
\end{CJK*}
\end{document}